\documentclass[letterpaper, twocolumn, 10pt]{article}
\usepackage[top=2.5cm,left=1.5cm,right=1.5cm,bottom=3cm]{geometry}

\usepackage{authblk}

\usepackage{amsmath,amssymb,amsfonts}
\usepackage{booktabs}
\usepackage{soul}
\usepackage{algorithm}
\usepackage{algorithmicx}
\usepackage[noend]{algpseudocode}
\usepackage{hyperref}
\usepackage{enumitem}
\usepackage{graphicx}
\graphicspath{{./images/}}
\usepackage{float}

\usepackage{comment}

\usepackage{makecell}
\usepackage{multirow}
\usepackage{booktabs}

\usepackage{textcomp}

\usepackage[svgnames]{xcolor}

\usepackage[breakable]{tcolorbox}

\usepackage{xspace}

\newcommand{\ra}[1]{\renewcommand{\arraystretch}{#1}}

\newcommand{\mT}{\tau_{\textsc{t}}}
\newcommand{\T}{$\mT$\xspace}
\newcommand{\mU}{\tau_{\textsc{u}}}
\newcommand{\U}{$\mU$\xspace}

\newtcolorbox{mybox}{
	arc=0pt,
	boxrule=0pt,
	colback=Gainsboro,
	width=\columnwidth,   
	colupper=black
}

\definecolor{babyblueeyes}{rgb}{0.63, 0.79, 0.95}

\newtcolorbox{mybox2}{
	arc=0pt,
	boxrule=0pt,
	breakable,
	colback=babyblueeyes,
	width=\columnwidth,   
	colupper=black
}

\begin{document}

\title{\Large \bf Adversarial Attacks against Binary Similarity Systems}

\author{{\rm Gianluca Capozzi}}

\author{{\rm Daniele Cono D'Elia}}

\author{{\rm Giuseppe Antonio Di Luna}}

\author{{\rm Leonardo Querzoni} \thanks{\texttt{\{capozzi, delia, diluna, querzoni\}@diag.uniroma1.it}}}

\affil[]{Sapienza University of Rome}

\date{}

\maketitle

\subsection*{Abstract}
In recent years, binary analysis gained traction as a fundamental approach to inspect software and guarantee its security. Due to the exponential increase of devices running software, much research is now moving towards new autonomous solutions based on deep learning models, as they have been showing state-of-the-art performances in solving binary analysis problems. One of the hot topics in this context is binary similarity, which consists in determining if two functions in assembly code are compiled from the same source code. However, it is unclear how deep learning models for binary similarity behave in an adversarial context.

In this paper, we study the resilience of binary similarity models against adversarial examples, showing that they are susceptible to both \textit{targeted} and \textit{untargeted} attacks (w.r.t. similarity goals) performed by {black-box} and {white-box} attackers.
In more detail, we extensively test three current state-of-the-art solutions for binary similarity against two black-box greedy attacks, including a new technique that we call \textit{Spatial Greedy}, and one white-box attack in which we repurpose a gradient-guided strategy used in attacks to image classifiers.

\section{Introduction}\label{sec:introduction}

An interesting problem that currently is a hot topic in the security and software engineering research communities~\cite{dullien2005graph, khoo2013rendezvous, alrabaee2015sigma}, is the {\em binary similarity problem}. That is to determine if two functions in assembly code are compiled from the same source code~\cite{massarelli2021function}. In this case, the two functions are {\em similar}. This problem is far from trivial: it is well-known that different compilers and optimization levels radically change the shape of the generated assembly code.

Binary similarity has many applications, including plagiarism detection, malware detection and classification, and vulnerability detection~\cite{DBLP:conf/pldi/DavidPY16, DBLP:conf/pldi/DavidPY17, egele2014blanket}. It can also be a valid aid for a reverse engineer as it helps with the identification of functions taken from well-known libraries or open-source software. Recent research~\cite{massarelli2021function} shows that techniques for binary similarity generalize, as they are are able to find similarities between semantically similar functions.

We can distinguish binary similarity solutions between the ones that use deep neural networks (DNNs), like~\cite{ding2019asm2vec, xu2017neural, massarelli2021function}, and the ones that do not, like~\cite{dullien2005graph, DBLP:conf/acsac/PewnySBHR14, DBLP:conf/pldi/DavidY14}.
Nearly all of the most recent works rely on DNNs, which offer in practice state-of-the-art performance while being computationally inexpensive. This aspect is particularly apparent when compared with solutions that build on symbolic execution or other computationally intensive techniques. 

However, a drawback of DNN-based solutions is their sensitivity to adversarial attacks~\cite{yuan2019adversarial} where an adversary crafts an innocuously looking instance with the purpose of misleading the target neural network model. Successful adversarial attacks have been well-documented for DNNs that process, for example, images~\cite{DBLP:journals/corr/SzegedyZSBEGF13, goodfellow2014explaining, carlini2017towards}, audio and video samples~\cite{DBLP:conf/nips/0001QLSKM19}, and text~\cite{DBLP:conf/emnlp/JiaL17}.

Binary similarity systems are an attractive target for an adversary. As examples, an attacker:
(1) may hide a malicious function inside a firmware by making it similar to a benign white-listed function, as similarly done in malware misclassification attacks~\cite{pierazzi2020intriguing};
(2) may make a plagiarized function dissimilar to the original one, analogously to source code authorship attribution attacks~\cite{DBLP:journals/pacmpl/Devore-McDonald20}; or, we envision, (3) may replace a function---entirely or partially, as in forward porting of bugs~\cite{DBLP:journals/pomacs/HazimehHP20}---with an old version known to have a vulnerability and make the result dissimilar from the latter.
	
In this paper, we say an attack is {\bf targeted} when the goal is to make a rogue function be the most similar to a target, as with example (1). An attack is {\bf untargeted} when the goal is to make a rogue function the most dissimilar from its original self, as with examples (2) and (3). In both contexts, the adversarial instance has to preserve the semantics of the rogue function as in its original form: that is, it must have the same execution behavior.

In spite of the wealth of works identifying similar functions with ever improving accuracy, we found that an extensive study on the resilience of (DNN-based) binary similarity solutions against adversarial attacks is missing.
	
In this paper, we aim to close this gap by proposing and evaluating techniques for targeted and untargeted attacks using both 
\textit{black-box} (where adversaries have access to the similarity model without knowing its internals) and \textit{white-box} (where they know also its internals) methods.

For the black-box scenario, we study a greedy approach that modifies a function by adding a single assembly instruction to its body at each optimization step. Where applicable, we also consider an enhanced gray-box~\cite{pierazzi2020intriguing} variant that, leveraging limited knowledge of the model, chooses only between instructions that the model treats as distinct. We then propose a novel black-box technique, which we call {\em Spatial Greedy}, where we transform the discrete space of assembly instructions into a continuous space using a technique based on instruction embeddings~\cite{DBLP:conf/nips/MikolovSCCD13}. This technique is on par or outperforms the gray-box greedy attack without requiring any knowledge of the model.
For the white-box scenario, we repurpose a method for adversarial attacks on images that relies on gradient descent~\cite{madry2017towards} and use it to drive instruction insertion decisions..

We test our techniques against three binary similarity systems---Gemini~\cite{xu2017neural}, GMN~\cite{DBLP:conf/icml/LiGDVK19}, and SAFE~\cite{massarelli2021function}---showing that they are vulnerable to targeted and untargeted attacks performed with our methods. For the best attack technique, the percentage of tested instances that could mislead the target model was 36.6\% for Gemini, 59.68\% for GMN, and 83.43\% for SAFE in the \textbf{targeted} scenario, while in the \textbf{untargeted} one it was, respectively, 53.89\%, 93.81\%, and 90.62\%. All three models appear inherently weaker in the face of an attacker seeking to stage untargeted attacks: with only a handful of added instructions, the attacker may evade state-of-the-art binary similarity models with high probability.

\subsection{Contributions}
This paper makes the following contributions:
\begin{itemize}
	\item we propose to study the problem of adversarial attacks against binary similarity systems, identifying targeted and untargeted attack opportunities;
	\item	we investigate black-box attacks against DNN-based binary similarity systems, exploring a greedy approach based on instruction insertion. Where applicable, we enhance it with partial knowledge of the model sensitivity to instruction types for efficiency;
	\item we propose Spatial Greedy, a fully black-box technique that matches or outperforms gray-box greedy by using embeddings to guide instruction selection;
	\item we investigate white-box attacks against DNN-based binary similarity systems, exploring a gradient-guided search strategy for inserting instructions;
	\item we conduct an extensive experimental evaluation of our techniques for targeted and untargeted attack scenarios against three systems backed by largely different models and with high performance in recent studies~\cite{marcelli2022machine}.
\end{itemize}


\section{Background}

In this section, we provide background knowledge for adversarial attacks against models for code analysis. Then, we introduce a categorization of semantics-preserving perturbations for binary functions.

\subsection{Adversarial Knowledge}\label{sec:NNThreats}
We can describe a deep learning model through different aspects: training data, layers architecture, loss function, and weights parameters. Having complete or partial knowledge about such elements can facilitate an attack from a computational point of view. According to seminal works in the area~\cite{BIGGIO2018317, pierazzi2020intriguing}, we can distinguish between:
\begin{itemize}
	\item \textbf{white-box} attacks, where the attacker has \textit{perfect knowledge} of the target model, including all the dimensions mentioned before. These type of attacks are realistic when the adversary has direct access to the model (e.g., an open-source malware classifier);
	\item \textbf{gray-box} attacks, where the attacker has \textit{partial knowledge} of the target model. For example, they have knowledge about feature representation (e.g., categories of features relevant for feature extraction);
	\item \textbf{black-box} attacks: the attacker has \textit{zero knowledge} of the target model. Specifically, the attacker is only aware of the task the model was designed for and has a rough idea of what potential perturbations to apply to cause some feature changes~\cite{BIGGIO2018317}.
\end{itemize}

Different attack types may suit different scenarios best. A white-box attack, for example, could be attempted on an open-source malware classifier. Conversely, a black-box attack would suit also a model hosted on a remote server to interrogate, as with a commercial cloud-based antivirus.

\begin{figure}[t!]
	\centering
	\includegraphics[width=2.95in,trim = 0cm 1cm 0cm 0cm]{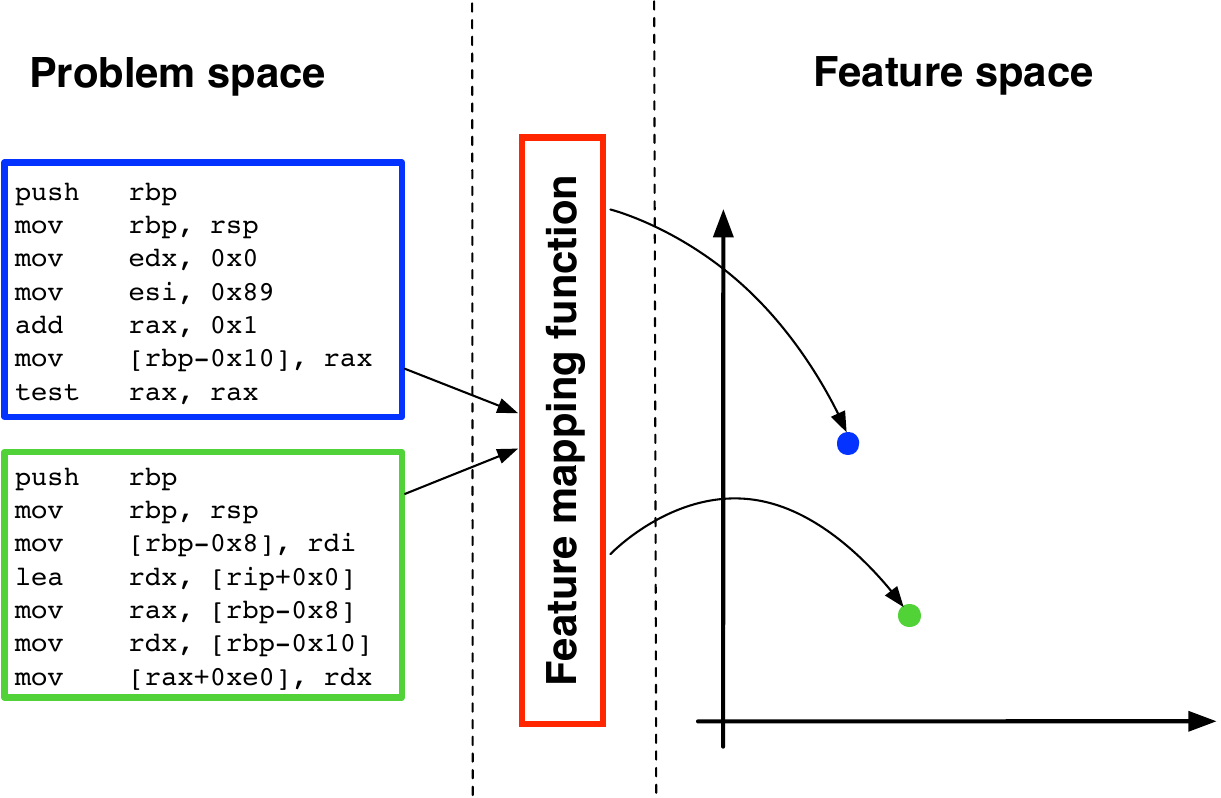}
	\caption{A feature mapping function maps problem-space objects into feature vectors. The two boxed binary functions implement similar functionalities, so they are mapped to two close points in the feature space.} 
	\label{img:featureMapping}
\end{figure}

\subsection{Inverse Feature Mapping Problem}\label{sec:invMapping}

In the following, we refer to the input domain as \textit{problem space} and to all its instances as \textit{problem-space} objects.

Deep learning models can manipulate only continuous {problem-space} objects. When inputs have a discrete representation, a first phase must map them into continuous instances. The phase usually relies on a \textit{feature mapping function} (Figure \ref{img:featureMapping}) whose outputs are \textit{feature vectors}. The set of all possible {feature vectors} is known as the \textit{feature space}.

Traditional white-box attacks against deep learning models solve an optimization problem in the {feature space} by minimizing an objective function in the direction following its negative gradient~\cite{madry2017towards}. When optimization ends, they obtain a {feature vector} that corresponds to a {problem-space} object representing the generated adversarial sample.

Unfortunately, given a {feature vector}, it is not always possible to obtain its {problem-space} representation. This issue is called the \textbf{inverse feature mapping} problem \cite{pierazzi2020intriguing}.

For {code models}, the feature mapping function is neither invertible nor differentiable. Therefore, one cannot understand how to modify an original {problem-space} object to obtain the given {feature vector}.
In particular, the attacker has to employ approximation techniques that create a feasible {problem-space} object from a feature vector. 
Ultimately, mounting an attack requires a manipulation of a {problem-space} object via perturbations guided by either  gradient-space attacks (as in the white-box case above) or ``gradient-free'' optimization techniques (as with black-box attacks). We discuss perturbations specific to our context next.

\subsection{Semantics-Preserving Perturbations of Problem-Space Objects}\label{sec:SemPres}

In this section, we discuss how to manipulate problem-space objects in the specific case of binary code models working on functions. To this end, we review and extend perturbations from prior works~\cite{pierazzi2020intriguing,DBLP:conf/asiaccs/SongLAGKY22,lucas2021malware}, identifying those suitable for adversarial manipulation of functions.

\begin{figure}[h]
	\centering
	\includegraphics[width=3.5in, trim = 0cm 0.5cm 0cm 0cm]{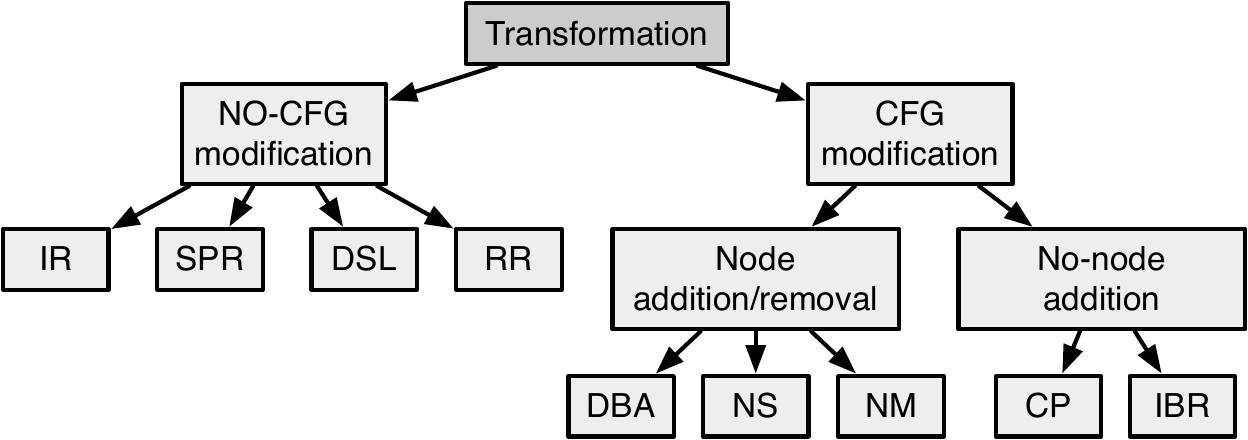}
	\caption{Taxonomy of semantics-preserving perturbations suitable for the proposed attacks. Acronyms are spelled out in the body of the paper.}
	\label{img:transformations}
\end{figure}

For our purpose, we seek to transform an original binary function $f$ into an adversarial binary sample $f_{adv}$ that preserves the semantics of $f$; intuitively, this restricts the set of available perturbations for the adversary. We report a taxonomy of possible \textit{semantics-preserving} perturbations in Figure \ref{img:transformations}, dividing them according to how they affect the binary layout of the function's control-flow graph (CFG). 

Among CFG-preserving perturbations, we identify:
\begin{itemize}
	\item \textbf{(IR) Instruction Reordering}: reorder independent instructions in the function;
	\item \textbf{(SPR) Semantics-Preserving Rewriting}: substitute a sequence of instructions with a semantically equivalent sequence;
	\item \textbf{(DSL) Modify the Data-Section Layout}: modify the memory layout of the \texttt{.data} section and update all the global memory offsets referenced by instructions;
	\item \textbf{(RR) Register Renaming}: change all the occurrences of a register as instruction operand with a register currently not in use or swap the use of two registers.
\end{itemize}

\begin{figure}[h]
	\centering
	\includegraphics[width=2.7in, trim = 0cm 0.5cm 0cm 0cm]{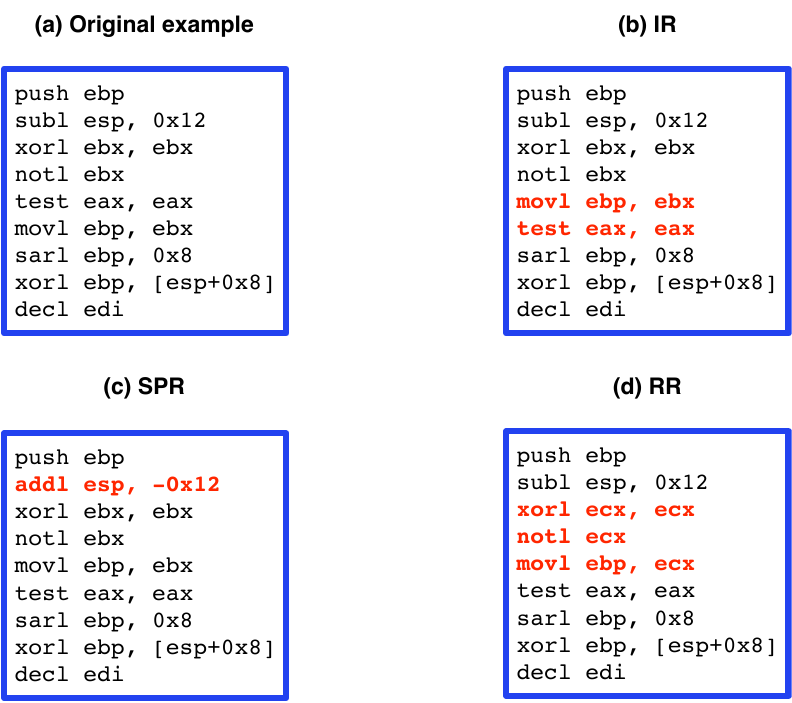}
	\caption{Examples of semantics-preserving perturbations that do not alter the binary CFG layout. We modify the assembly snippet in \textbf{(a)} by applying, in turn, \textbf{(b)} Instruction Reordering, \textbf{(c)} Semantics-Preserving Rewriting, and \textbf{(d)} Register Renaming. Altered instructions are in \textcolor{red}{\textbf{red}}.}
	\label{img:noCFGtransformations}
\end{figure}

Figure~\ref{img:noCFGtransformations} shows examples of their application.
As for perturbations that affect the (binary-level) CFG layout, we can identify the ones that involve adding or deleting nodes:
\begin{itemize}
	\item  \textbf{(DBA) Dead Branch Addition}: add dead code in a basic block guarded by an always-false branch;
	\item  \textbf{(NS) Node Split}: split a basic block without altering the semantics of its instructions (e.g., the original block will jump to the one introduced with the split);
	\item   \textbf{(NM) Node Merge}: merge two basic blocks when semantics can be preserved. For example, by using predicated execution to linearize branch-dependent assignments as conditional {\tt mov} instructions~\cite{constantine}. 
\end{itemize}

And the ones that leave the graph structure unaltered:
\begin{itemize}
	\item  \textbf{(CP) Complement Predicates}: change the predicate of a conditional branch and the branch instruction with their negated version;
	\item  \textbf{(IBR) Independent Blocks Reordering}: change the order in which independent basic blocks appear in the binary representation of the function.
\end{itemize}

\section{Threat Model and Problem Definition}
In this section, we define our threat model together with the problem of attacking binary similarity models.

\subsection{Threat Model}
\label{ss:threat-model}
The focus of this work is to create adversarial instances that attack a model at inference time (i.e., we do not investigate attacks at training time). Following the description provided in Section \ref{sec:NNThreats}, we consider two different attack scenarios: respectively, a black-box and a white-box one. In the first case, the adversary has no knowledge of the target binary similarity model; nevertheless, they can perform an unlimited number of queries to observe the output produced by the model. In the second case, we assume that the attacker has perfect knowledge of the target binary similarity model. 

\subsection{Problem Definition}
\label{ss:problem-definition}
Let $sim$ be a similarity function that takes as input two functions, $f_1$ and $f_2$, and returns a real number, the {\em similarity score} between them, in $[0,1]$. 

\footnotetext[1]{Although $f_{adv}$ and $f_2$ are similar for the model, they are not semantically equivalent: this is precisely the purpose of an attack that wants to fool the model to consider them as such, while they are not.}

We define two binary functions to be \textit{semantically equivalent} if they are two implementations of the same abstract functionality. We assume that there exists an adversary that wants to attack the similarity function. The adversary can mount two different kind of attacks:
\begin{itemize}[noitemsep,topsep=2pt]
	\item \textbf{Targeted attack}. Given two binary functions, $f_1$ (identified as \textit{source}) and $f_2$ (identified as \textit{target}), the adversary wants to find a binary function $f_{adv}$  semantically equivalent to $f_1$ such that: $sim(f_{adv}, f_{2}) \ge \mT$, where $\mT$ is a success threshold\footnotemark{} chosen by the attacker depending on the victim at hand.
	\item \textbf{Untargeted attack}. Given a binary function $f_1$, the adversary goal consists of finding a binary function $f_{adv}$ semantically equivalent to $f_1$ such that: $sim(f_{1}, f_{adv}) \le \mU$. The threshold $\mU$ is the analogous of the previous case for the untargeted attack scenario.
\end{itemize}
Loosely speaking, in the first case, the adversarial sample has to be similar to a target, as in the example scenario (1) presented in Section \ref{sec:introduction}. In the second one, the adversarial sample has to be dissimilar from its original version, as in the example scenarios (2) and (3) also from Section \ref{sec:introduction}. 

\begin{figure*}[t!]
	\centering
	\includegraphics[width=7in, trim = 0cm 1.2cm 0cm 0cm]{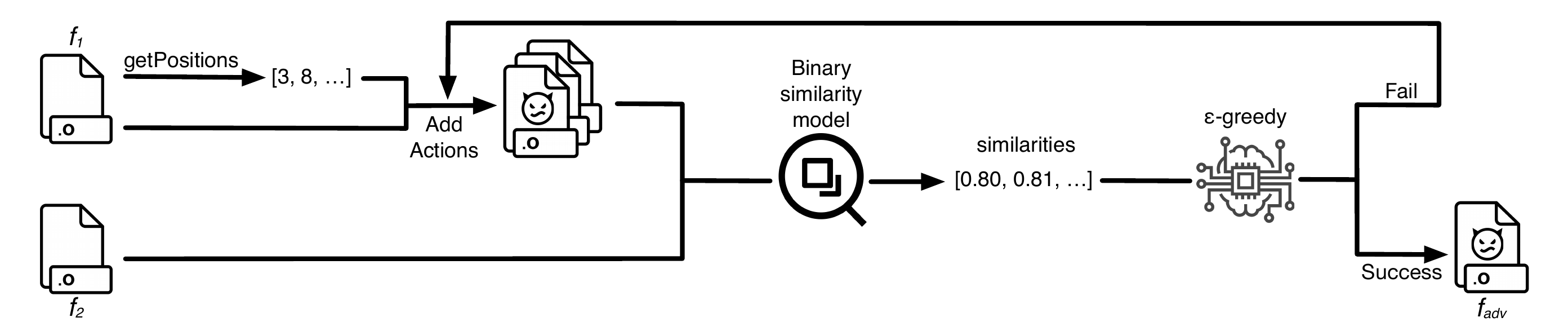}
	\caption{Overall workflow of the \textit{black-box} $\varepsilon$-greedy action-selection strategy in the \textit{targeted} scenario.}
	\label{img:advWork}
\end{figure*}

\subsection{Perturbation Selection}\label{ss:perturbation-selection}
Given a binary function $f_1$, our attack consists in applying to it perturbations that do not alter its semantics.

To study the feasibility of our approach, we choose \textit{dead branch addition} (DBA) among the suitable perturbations outlined in Section \ref{sec:SemPres}. We find DBA adequate for this study for two reasons: it is sufficiently expressive so as to affect heterogeneous models (which may not hold for others\footnotemark{}) and its implementation complexity for an attacker is fairly limited. Nonetheless, other choices remain possible, as we will further discuss in Section~\ref{se:limitations}.

At each application, our embodiment of DBA inserts in the binary code of $f_1$ one or more instructions in a new or existing basic block guarded by a branch that is never taken at runtime (i.e., we use an always-false branch predicate).

Such a perturbation can be done at compilation time or on an existing binary function instance. For our study, we apply DBA during compilation by adding placeholder blocks as inline assembly, which eases the generation of many adversarial samples from a single attacker-controlled code. State-of-the-art binary rewriting techniques would work analogously over already-compiled source functions.

We currently do not attempt to conceal the nature of our branch predicates for preprocessing robustness, which~\cite{pierazzi2020intriguing} discusses as something that attackers should be wary of to mount stronger attacks. We believe off-the-shelf obfuscations (e.g., opaque predicates, mixed boolean-arithmetic expressions) or more complex perturbation choices may improve our approach in this respect. Nevertheless, our main goal was to investigate its feasibility in the first place.


\section{Black-Box attack: Solution Overview}\label{sec:blackbox}
In this section, we describe our black-box attacks. We first introduce our baseline solution (named \textbf{Greedy}), highlighting its limitations. We then move to our main contribution in the black-box scenario (named \textbf{Spatial Greedy}). Figure \ref{img:advWork} depicts a general overview of our black-box approaches.

\subsection{Greedy}\label{sec:greedy}
The baseline black-box approach we devise for attacking binary function similarity models consists of an iterative action-selection rule that follows a greedy optimization strategy. Starting from the original sample $f_{1}$, we iteratively apply perturbations $T_1, T_2, \ldots, T_k$ selected from a set of available actions, generating a series of samples $f_{adv_1}, f_{adv_2}, \ldots, f_{adv_k}$. This procedure ends upon generating a sample $f_{adv}$ meeting the desired similarity threshold, otherwise the attack fails after $\bar{\delta}$ completed iterations.

\footnotetext[2]{For example, basic block-local transformations such as IR and RR would have limited efficacy on models that study an individual block for its instruction types and counts or other coarse-grained abstractions. This is the case with Gemini and GMN that we attack in this paper.}

For instantiating Greedy using DBA perturbations, we reason on a set of positions $\texttt{BLK}$ for inserting dead branches in function $f_{1}$ and a set of instructions $\texttt{CAND}$, which we call the \textit{set of candidates}. Each action consists of a $\langle \texttt{bl},\texttt{in} \rangle$ pair made of  the branch $\texttt{bl} \in \texttt{BLK}$ and an instruction $\texttt{in} \in \texttt{CAND}$ to insert in the dead code block guarded by $\texttt{bl}$.

The naive action-selection rule (i.e., greedy) at each step selects the action (i.e., the perturbation) that locally maximizes (or minimizes in case of untargeted attack) the relative increase (or decrease) of the objective function.

This approach, however, may be prone to finding local optima. To avoid this problem, we choose as our Greedy baseline an $\varepsilon$-greedy action-selection rule. Here, we select with a small probability $\varepsilon$ a suboptimal action instead of the one that the standard greedy strategy picks, and with probability $1-\varepsilon$ the one representing the local optimum.

The objective function is the similarity between $f_{adv}$ and the target function $f_{2}$ in case of a targeted attack (formally, $sim(f_{adv}, f_2)$) or the negative of the similarity between $f_{adv}$ and the original function in case of untargeted attack (formally, $-sim(f_{1}, f_{adv})$). In the following, we only discuss the maximization strategy followed by targeted attacks; mutatis mutandis, the same rationale holds for untargeted attacks.

\subsubsection{Limitations of the Complete Enumeration Strategy}
At each step, Greedy enumerates all the applicable transformations computing the marginal increase of the objective function, thus resulting in selecting an instruction $\texttt{in}$ by enumerating all the possible instructions of the considered set of candidates $\texttt{CAND}$ for each position $\texttt{bl} \in \texttt{BLK}$.

Unfortunately, the instruction set architecture (ISA) of a modern CPU may consist of a large number of instructions. To give an example, consider the x86-64 ISA: according to~\cite{DBLP:conf/pldi/HeuleS0A16}, it has 981 unique mnemonics and a total of 3,684 instruction variants (without counting register operand choices for them). 
Therefore, it would be unfeasible to have a $\texttt{CAND}$ set that covers all possible instructions of an ISA.

This means that the size of $\texttt{CAND}$ must be limited. One possibility is to use hand-picked instructions. However, this approach has two problems. Such a set could not cover all the possible behaviors of the ISA, missing fundamental aspects (for example, leaving vector instructions uncovered); furthermore, this effort has to be redone for a new ISA. There is also a more subtle pitfall: a set of candidates fixed in advance could include instructions that the specific binary similarity model under attack deems as not significant.

On specific models, it may still be possible to use a small set of candidates profitably, enabling a \textbf{gray-box} attack strategy for Greedy. In particular, one can restrict the set of instructions to the ones that effectively impact the features extracted by the attacked model (which obviously requires knowledge of the features it uses; hence, the gray-box characterization). In such cases, this strategy is equivalent to the black-box Greedy attack that picks from all the instructions in the ISA, but computationally much more efficient.

\subsection{Spatial Greedy}\label{sec:spatial}

In this section, we extend our baseline solution by introducing a fully black-box approach named Spatial Greedy. This approach overcomes all the limitations discussed for Greedy using an adaptive procedure that dynamically updates the set of candidates according to a feedback from the model under attack \textit{without requiring any knowledge} of it. 

In Spatial Greedy, we extend the $\varepsilon$-greedy action-selection strategy by adaptively modifying the set of candidates that we use at each iteration. In particular, using instructions embedding techniques, we transform each instruction $\texttt{in} \in \texttt{CAND}$ into a vector of real values. This creates vectors that partially preserve the semantics of the original instructions. Chua et al.~\cite{DBLP:conf/uss/ChuaSSL17} first showed that such vectors may be grouped by instruction semantics, creating a notion of proximity between instructions: for example, vectors representing arithmetic instructions are in a cluster, vectors representing branches in another, and so on.

Here, at each step, we populate a portion of the set of candidates by selecting the instructions that are close, in the embedding metric space, to instructions that have shown a good impact on the objective function. The remaining portion of the set is composed of random instructions. We discuss our choices for instruction embedding techniques and dynamic candidates selection in the following.

In the experimental section, for the black-box realm, we will compare Spatial Greedy against the Greedy approach, opting for the computationally efficient gray-box flavor of the latter when allowed by the specific model under study.

\begin{figure}[t!]
\centering
\includegraphics[width=3.5in, trim = 0cm 0.8cm 0cm 0cm]{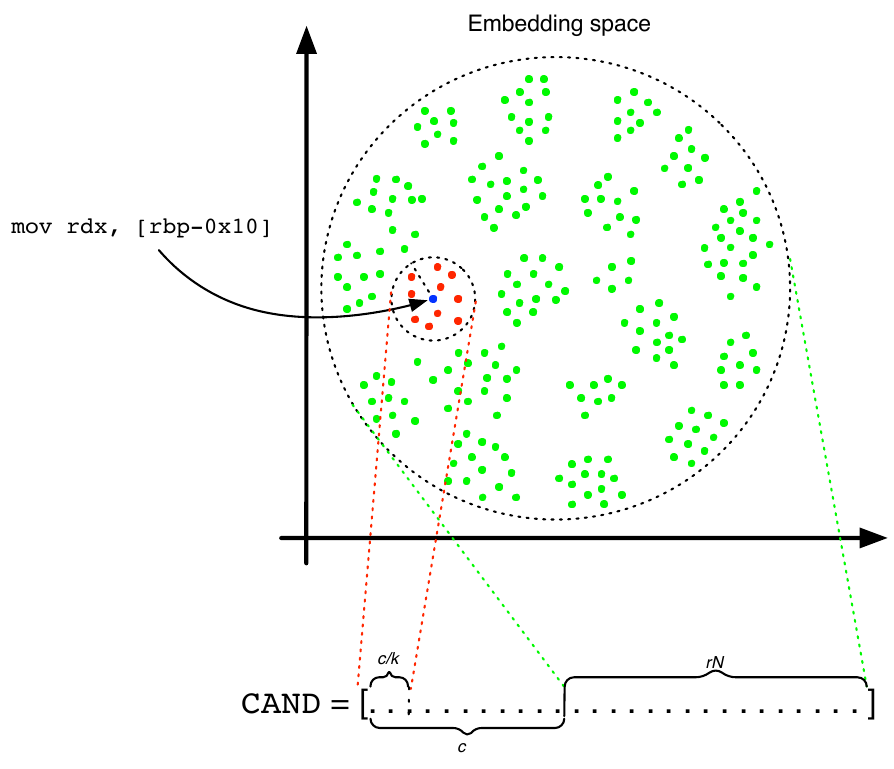}
\caption{Dynamic update of the set of candidates. The $\mathtt{mov}$ instruction is the greedy action for the current iteration and is mapped to the \textcolor{blue}{\textbf{blue}} point in the instruction embedding space. The set of candidates is updated selecting $c/k$ neighbours of the considered \textit{top-k} action (represented in \textcolor{red}{\textbf{red}}), $c-c/k$ instructions among the closest neighbours of the remaining \textit{top-k} greedy actions, and $rN$ random instructions.}
\label{img:spatial}
\end{figure}

\subsubsection{Instruction Embedding Space}\label{sec:proxy}
We embed assembly instructions into numeric vectors using an instruction embedding model~\cite{DBLP:conf/nips/MikolovSCCD13}. Given such a model $M$ and a set $I$ of assembly instructions, we map each $i \in I$ to a vector of real values $\vec{i} \in \mathbb{R}^n$, using $M$. The model is such that, for two instructions having similar semantics, the embeddings it produces will be close in the metric space.

\subsubsection{Dynamic Selection of the Set of Candidates}\label{sec:spatialgreedy:selection}
The process for selecting the set of candidates for each iteration of the $\varepsilon$-greedy action-selection procedure represents the focal point of Spatial Greedy.

Let $N$ be the size of the set of candidates $\texttt{CAND}$. Initially, we fill it with N random instructions. Then, at each iteration of the $\varepsilon$-greedy procedure, we update $\texttt{CAND}$ by inserting $rN$ random instructions, where $r \in [0,1)$, and $c$ instructions we select among the closest neighbors of the instructions composing the {top-$k$ greedy actions} of the last iteration.

The {\em top-$k$ greedy actions} are the $k$ actions that, at the end of the last iteration, achieved the highest increase of the objective function. To keep the size of the set stable at value $N$, we take the closest $c/k$ neighbors of each top-$k$ action\footnote{We also apply rounding so that we can work with integer numbers.}.

The rationale of having $r$ random and $c$ selected instructions is seeking a balance between {\em exploration} and {\em exploitation}. With the random instructions, we randomly sample the solution space to escape from a possibly local optimum found for the objective function. With the selected instructions, we exploit the part of the space that in the past has brought the best solutions. Figure \ref{img:spatial} provides a pictorial representation of the update procedure.

We present the complete description of Spatial Greedy in Algorithm \ref{spatGreedyPseudocode}. The first step consists in identifying the positions $\texttt{BLK}$ where to introduce dead branches and initializing the set of candidates $\texttt{CAND}$ with $N$ random instructions (lines \ref{alg:getPositions} and \ref{alg:initCandidates}). Then, during the iterative procedure (lines \ref{alg:startIters}-\ref{alg:endIters}), we first enumerate all the possible actions (lines \ref{alg:startActs}-\ref{alg:endActs}), then apply the action-selection rule according to the value of $\varepsilon$ (lines \ref{alg:startSel}-\ref{alg:endSel}). Finally, we get the \textit{top-k greedy actions} (line \ref{alg:getTopK}) and update the set of candidates (line \ref{alg:updateCandidates}).

\begin{algorithm}[t!]
\begin{small}
\footnotesize
\caption{Spatial Greedy procedure (\textbf{targeted} case)}
\label{spatGreedyPseudocode}
\textbf{Input}: source function $f_1$, target function $f_2$, similarity threshold \T, max number of dead branches $B$, max number of instructions to be inserted $\bar{\delta}$, max number of instructions to be tested $N$, max number of random instructions $r$, max number of neighbours $c$, probability of selecting random action $\varepsilon$. \textbf{Output}: adversarial sample $f_{adv}$.
\begin{algorithmic}[1]
	\State $f_{adv} \leftarrow f_1$
	\State $instr \leftarrow 0$
	\State $\texttt{BLK} \leftarrow \textrm{getPositions}(f_1, B)$ \label{alg:getPositions}
	\State $\texttt{CAND} \leftarrow \textrm{getRandomInstructions}(N)$ \label{alg:initCandidates}
	\State $sim \leftarrow \textrm{sim}(f_{adv}, f_{2})$
	\While{$sim \le $ \T AND $instr < \bar{\delta}$} \label{alg:startIters}
	\State $iterSim \leftarrow sim$
	\State $iterBlock \leftarrow \langle \rangle$
	\State $testedActions \leftarrow [\:]$
	\For{$\langle \texttt{bl},\texttt{in} \rangle \in \texttt{BLK} \times \texttt{CAND}$} \label{alg:startActs}
	\State $\overline{f}_{adv} \leftarrow f_{adv}+ \langle \texttt{bl},\texttt{in} \rangle$
	\State $currSim \leftarrow \textrm{sim}(\overline{f}_{adv}, f_2)$
	\State $testedActions.\textrm{append}(\langle \langle \texttt{bl},\texttt{in} \rangle, currSim \rangle)$
	\State $prob \leftarrow \textrm{uniform}(0,1)$
	\EndFor \label{alg:endActs}
	\If{$\textrm{prob} < \varepsilon$} \label{alg:startSel}
	\State $iterSim, iterBlock \leftarrow \textrm{selectGreedy}()$
	\Else
	\State $iterSim, iterBlock \leftarrow \textrm{selectRandom}()$
	\EndIf \label{alg:endSel}
	\State $elected \leftarrow \textrm{getTopK}(testedActions, K)$ \label{alg:getTopK}
	\State $\texttt{CAND} \leftarrow \textrm{updateInstructions}(elected, r, c)$ \label{alg:updateCandidates}
	\State $sim \leftarrow iterSim$
	\State $f_{adv} \leftarrow f_{adv} + iterBlock$
	\State $instr\leftarrow instr + 1$
	\EndWhile \label{alg:endIters}
	\State \Return $f_{1}'$
\end{algorithmic}
\end{small}
\end{algorithm}

\section{White-Box attack: Solution Overview} \label{sec:whitebox}
As pointed out in Section \ref{sec:NNThreats}, in a white-box scenario the attacker has a perfect knowledge of the target deep learning model, including its loss function and gradients. We discuss next how we can build on them to mount an attack.

\subsection{Gradient-guided Code Addition Method}\label{sec:GCAM}
White-box adversarial attacks have been largely investigated against image classifiers by the literature, resulting in valuable effectiveness~\cite{goodfellow2014explaining}. Our attack strategy for binary similarity derives from the design pattern of the PGD attack~\cite{madry2017towards}, which iteratively targets image classifiers.

We call our proposed white-box attack \textit{Gradient-guided Code Addition Method} (\textbf{GCAM}). It consists in applying a set of transformations using a gradient-guided strategy. In particular, we try to minimize (or maximize, depending on whether the attack is targeted or untargeted) the loss function of the attacked model on the given input while keeping the size of the perturbation small (and respecting the semantics-preserving constraint): specifically, we use for this purpose $L_p$-norm as soft constraint.

Because of the \textit{inverse feature mapping} problem, gradient optimization-based approaches cannot be directly applied in our context (Section~\ref{sec:invMapping}). We need a further (hard) constraint that acts on the feature-space representation of the input binary function. This constraint strictly depends on the target model: we will further investigate its definition in Section \ref{chap:targetSystems}. For the following, we focus on the loss minimization strategy argued for targeted attacks. As before, we can easily adapt the same concepts to the untargeted case.

We can describe a DNN-based model for binary similarity as the concatenation of the two functions $\lambda$ and $sim_v$. In particular, $\lambda$ is the function that maps a problem-space object to a feature vector (i.e., the feature mapping function discussed in Section \ref{sec:invMapping}), while $sim_v$ is the neural network computing the similarity given the feature vectors.

Given two binary functions $f_1$ and $f_2$, we aim to find a perturbation $\delta$ that maximizes $sim_v(\lambda(f_1))+\delta, \lambda(f_2))$. To do so, we use an iterative strategy where, during each iteration, we solve the following optimization problem:
\begin{equation} \label{eq:whiteAttack}
\begin{aligned}
& \text{min}
& & \mathcal{L}(sim_v(\lambda(f_1)+\delta, \lambda(f_2)),\theta)  + \epsilon ||\delta||_{p},\\
\end{aligned}
\end{equation}
where $\mathcal{L}$ is the loss function, $\theta$ are the weights of the target model, and $\epsilon$ is a coefficient in $[0,\infty)$.  

We randomly initialize the perturbation $\delta$ and then update it at each iteration by a quantity given by the negative gradient of the loss function $\mathcal{L}$. The vector $\delta$ has several components equal to zero and it is crafted so that it modifies only the (dead) instructions in the added blocks. The exact procedure depends on the target model: we return to this aspect in Section \ref{chap:targetSystems}.

Notice that the procedure above allows us to find a perturbation in the feature space, while our final goal is to find a problem-space perturbation to modify the function $f_1$. Therefore, we derive from the perturbation $\delta$ a problem-space perturbation $\delta_p$. The exact technique is specific to the model we are attacking, as we further discuss in Section \ref{chap:targetSystems}.

The common idea behind all technique instances is to find the problem-space perturbation $\delta_p$ whose representation in the feature space is the {\em closest} to $\delta$. Essentially, we use a {\em rounding-based inverse} strategy to solve the inverse feature mapping problem that accounts to {\em rounding} the feature space vector to the closest vector that corresponds to an object in the problem space.
The generated adversarial function is $f_{adv} = f_1 + \delta_p$. As for the black-box scenario, the process ends whenever we reach a maximum number of iterations or the desired threshold value for $sim$.


\section{Target systems}\label{chap:targetSystems}

In this section, we illustrate the three models we attacked: Gemini \cite{xu2017neural}, GMN \cite{DBLP:conf/icml/LiGDVK19}, and SAFE \cite{massarelli2021function}. We give a high-level description of their internals and then discuss specific provisions for the Greedy (Section \ref{sec:greedy}) and GCAM (Section \ref{sec:whitebox}) attacks---whereas Spatial Greedy needs no adaptations.

For their choice, we first surveyed recent comparative evaluations (most prominently~\cite{marcelli2022machine}) to identify plausible, performant candidates. We then reviewed the characteristics of the neural networks and the code analysis choices behind them, identifying three distinctly different approaches to problem solving. As we will see, for example, GMN and Gemini rely on CFG properties whereas SAFE does not; GMN accounts for graph similarity using an explicit matching mechanism that is different from the message passing network behind Gemini; SAFE employs a different kind of neural network than the other two systems. 

\subsection{Gemini}\label{sec:gemini}

Gemini \cite{xu2017neural} represents functions in the problem space through their Attributed Control Flow Graph (ACFG). An ACFG is a control flow graph where each basic block consists of a vector of manual features (i.e., node embeddings).

The focal point of this approach consists of a graph neural network (GNN) based on the Structure2vec~\cite{DBLP:conf/icml/DaiDS16} model that converts the ACFG into an embedding vector, obtained by aggregating the embedding vectors of individual ACFG nodes. The similarity score for two functions is given by the cosine similarity of their ACFG embedding vectors.

\subsubsection{Greedy Attack}\label{sec:black:gemini}
Each ACFG node contributes a vector of 8 manually selected features. Five of these features depend on the characteristics of the instructions in the node, while the others on the graph topology. The model distinguishes instructions from an ISA only for how they contribute to these 5 features.
This enables a gray-box variant of our Greedy attack: we measure the robustness of Gemini using a set of candidates $\texttt{CAND}$ of only five instructions, carefully selected for covering the five features. Later in the paper, we use this variant as the baseline approach for a comparison with Spatial Greedy.

\subsubsection{GCAM Attack}\label{sec:white:gemini}

As described in the previous section, some of the components of a node feature vector $v$ depend on the instructions inside the corresponding basic block. As Gemini maps all possible ISA instructions into 5 features, we can associate each instruction with a deterministic modification of $v$ represented as a vector $u$. We select five categories of instructions and for each category $c_j$ we compute the modification $u_j$ that will be applied to the feature vector $v$. We selected the categories so as to cover the aforementioned features.

When we introduce in the block an instruction belonging to category $c_j$, we add its corresponding $u_j$ modification to the feature vector $v$. Therefore, adding instructions inside the block modifies the feature vector $v$ by adding to it a linear combination vector $\sum_{j}n_j u_j$, where $n_j$ is the number of instructions of category $c_j$ added. Our perturbation $\delta$ acts on the feature vector of the function only in the components corresponding to the added dead branches, by modifying the coefficients of the linear combination above.

Since negative coefficients are meaningless, we avoid them by adding to the optimization problem appropriate constraints. Moreover, we solve the optimization problem without forcing the components of $\delta$ to be integers, as this would create an integer programming problem. Therefore, at the end of the iterative optimization process, we get our problem-space perturbation $\delta_p$ by \textit{rounding} each component of $\delta$. It is immediate to obtain from $\delta_p$ the problem-space modification to our binary function $f_1$. Indeed, in each dead block, we must add as many instructions belonging to a category as the corresponding coefficient in $\delta_p$.

\subsection{GMN}\label{sec:gmn}
Graph Matching Network (GMN)~\cite{DBLP:conf/icml/LiGDVK19} computes the similarity between two graph structures. When functions are represented through their CFGs, GMN offers state-of-the-art performance for the binary similarity problem~\cite{DBLP:conf/icml/LiGDVK19,marcelli2022machine}.

Differently from solutions based on standard GNNs (e.g., Gemini), which compare embeddings built separately for each graph, GMN computes the distance between two graphs as it attempts to match them. In particular, while in a standard GNN the embedding vector for a node captures properties of its neighborhood only, GMN also accounts for the similarity with nodes from the other graph.

\subsubsection{Greedy Attack}\label{sec:black:gmn}
Similarly to the case of Gemini, each node of the graph consists of a vector of manually-engineered features. In particular, each node is a bag of 200 elements, each of which represents a class of assembly instructions, grouped according to their mnemonics. The authors do not specify why they only consider these mnemonics among all the available ones in the \texttt{x86-64} ISA.
Analogously to Gemini, when testing the robustness of this model against the Greedy approach we devise a gray-box variant by considering a set of candidates $\texttt{CAND}$ of 200 instructions, each of which belonging to one and only one of the considered classes.

\subsubsection{GCAM Attack}\label{sec:white:gmn}
Our white-box attack operates analogously to what we presented in Section \ref{sec:white:gemini}.
Similarly to the Gemini case, each dead branch adds a node to the CFG while the feature mapping function transforms each CFG node into a feature vector. The feature vector is a bag of the instructions contained in the node, where assembly instructions are divided into one of 200 categories using the mnemonics.

\subsection{SAFE}\label{sec:safe}
SAFE~\cite{massarelli2021function} is an embedding-based similarity model. It represents functions in the problem space as sequences of assembly instructions.
It first converts assembly instructions into continuous vectors using an instruction embedding model based on the word2vec~\cite{DBLP:conf/nips/MikolovSCCD13} word embedding technique.
Then, it supplies such vectors to a bidirectional self-attentive recurrent neural network (RNN), obtaining an embedding vector for the function. The similarity between two functions is the cosine similarity of their embedding vectors.

\subsubsection{Greedy Attack}\label{sec:black:safe}
The Greedy attack against SAFE follows the black-box approach described in Section \ref{sec:greedy}. Since SAFE does not use manually engineered features, we cannot select a restricted set of instructions that generates all vectors of the feature space for a gray-box variant. We test its resilience against the Greedy approach considering a carefully designed list of candidates $\texttt{CAND}$ composed of random and hand-picked instructions, meaning that the baseline is a black-box attack.

\subsubsection{GCAM Attack}\label{sec:white:safe}
In the feature space, we represent a binary function as a sequence of instruction embeddings belonging to a predefined metric space. The perturbation $\delta$ is a sequence of real-valued vectors initialized with embeddings of real random instructions; each dead block contains four of such vectors. In the optimization process, we modify each embedding $i_j \in \delta$ by a small quantity given by the negative gradient of the loss function $\mathcal{L}$. In other words, every time we optimize the objective function, we alter each $i_j \in \delta$ by moving it in the negative direction identified through the gradient.

Since during optimization we modify instruction embeddings in terms of their single components, we have no guarantee that the obtained vectors are embeddings of real instructions. For this reason, after the optimization process, we compute the problem-space perturbation $\delta_p$ by {\em rounding}, at each iteration, the embeddings in $\delta$ to the closest embeddings in the space of real instruction embeddings.


\section{Datasets and Implementation}
In this section, we discuss the evaluation datasets and the corpus for training the embedding model of Spatial Greedy.

\subsection{Attack Dataset}\label{sec:dataset}

We test our approaches by considering pairs of binary functions randomly extracted from 6 open-source projects written in C language: binutils, curl, gsl, libconfig, libhttp, and openssl.
We compile the programs for an \texttt{x86-64} architecture using the \texttt{gcc} 9.2.1 compiler with \texttt{-O0} optimization level on Ubuntu 20.04. We filter out all functions with less than six instructions.
As a result, we obtain a dataset of code representative of real-world software, with source programs used in the evaluation or training of binary similarity solutions (e.g.,~\cite{marcelli2022machine,massarelli2021function,ding2019asm2vec,xu2017neural}), and that could be potential targets for the exemplary scenarios outlined in Section~\ref{sec:introduction}.

To evaluate the robustness of the three target models against our proposed approaches, we used two datasets for the \textbf{targeted} scenario, each made of 500 pairs of binary functions sampled from the general dataset. For both datasets, the same function cannot be considered twice as source function, but it can appear as target more than once. The first dataset, which we call \textsf{Random}, consists of pairs of random functions: functions in a pair differ at most by 1345 instructions, and on average by 135.27. The second dataset, which we call \textsf{Balanced}, includes pairs of functions having similar length. In particular, they differ at most by 10 instructions, and on average by 5.47. We make no attempt to balance either dataset by considering the number of CFG nodes: for them, we measure an average difference of 17.8 nodes in \textsf{Random} and 7.5 in \textsf{Balanced}.

In the \textbf{untargeted} scenario, source and target functions have to coincide. For these attacks, we use the dataset \textsf{Untarg} composed by the 500 functions used as source in the \textsf{Random} dataset. Being pairs made of identical functions, they are trivially balanced for instructions and CFG nodes.

\subsection{Dataset used for Spatial Greedy}\label{sec:SG_dataset}
As described in Section~\ref{sec:proxy}, in Spatial Greedy we use an instruction embedding model to induce a metric space over assembly instructions. We opt for word2vec~\cite{DBLP:conf/nips/MikolovSCCD13}, being it currently the state-of-the-art solution in the field\footnotemark{}. For each of the considered models, we use the following parameters during training: embedding size 100, window size 8, word frequency 8 and learning rate 0.05.
We train these models using assembly instructions as tokens. We use as training set a corpus of {23,181,478} assembly instructions, extracted from {291,688} binary functions collected by compiling various system libraries with the same setup of the previous section.

One aspect worth emphasizing is that Spatial Greedy uses embeddings unrelated to the binary similarity model being targeted. We trained the Spatial Greedy embedding model using distinct dataset and parameters compared to SAFE, whereas neither GMN nor Gemini incorporate a layer that converts a single instruction into a feature vector. Spatial Greedy relies on embeddings to enhance instruction selection during the attack by clustering the instruction space, independently of the underlying model being attacked.

\subsection{Implementation details}
We implement our attacks in Python in about 3100 LOC.

An aspect that is worth mentioning for the greedy attacks involves the application of the action $\langle \texttt{bl}, \texttt{in} \rangle$ chosen at each iteration. We note that modifying the binary representation of the function every time incurs costs (recompilation in our case; binary rewriting in alternative implementations) that we may avoid through a simulation.

In particular, we directly modify the data structures that the target models use for feature mapping when parsing the binary, simulating the presence of newly inserted instructions. These models have been implemented by their authors in tensorflow or pytorch, which allows us to keep our modifications rather limited. In preliminary experiments, we have verified that the similarity values from our simulation are comparable with those we would have obtained by recompiling the modified functions output by our attacks. Hence, we will use it for the experiments of Section~\ref{sec:Eval}.

\footnotetext{The attentive reader may wonder whether this choice may unfairly favor Spatial Greedy when attacking SAFE, as the model also uses word2vec in its initial instruction embedding stage. We conducted additional experiments for SAFE using two other models, GloVe~\cite{pennington2014glove} and fastText~\cite{bojanowski2016enriching}. The three models perform almost identically in targeted attacks, while in untargeted ones fastText occasionally outperforms the others by a small margin. For the sake of generality, in the paper evaluation we will report and discuss results for word2vec only.}

\begin{table*}[t!]
	\footnotesize
	\ra{1.5}
	\caption{Evaluation metrics with $\mT = 0.80$ relative to the \textbf{black-box} attacks against the three target models in the \textbf{targeted} scenario. Spatial Greedy (SG) is evaluated using parameters $\varepsilon = 0.1$ and $r = 0.75$. Greedy (G) is evaluated using $\varepsilon = 0.1$. \textcolor{blue}{\textbf{G*}} is the gray-box version of Greedy: when such a version is available (Section~\ref{chap:targetSystems}), we show it instead of G. When examining G against SAFE, a set of candidates of size 400 is considered.}
	\label{tab:BBTargeted}
	\centering
	\resizebox{\linewidth}{!}{%
		\begin{tabular}{c|c|c|cccc|cccc|cccc|cccc} 
			\toprule
			\multicolumn{1}{l|}{\multirow{2}{*}{}} & \multicolumn{1}{l|}{\multirow{2}{*}{\textbf{Target}}} & \multicolumn{1}{l|}{\multirow{2}{*}{\textbf{Attack}}}                        & \multicolumn{4}{c|}{\textbf{A-rate (\%) ($\mT = 0.80$)}} & \multicolumn{4}{c|}{\textbf{M-size ($\mT = 0.80$)}}  & \multicolumn{4}{c|}{\textbf{A-sim ($\mT = 0.80$)}}   & \multicolumn{4}{c}{\textbf{N-inc ($\mT = 0.80$)}}     \\ 
			\cline{4-19}
			\multicolumn{1}{l|}{}                  & \multicolumn{1}{l|}{}                                 & \multicolumn{1}{l|}{}                                                        & \textbf{C1} & \textbf{C2} & \textbf{C3} & \textbf{C4}     & \textbf{C1} & \textbf{C2} & \textbf{C3} & \textbf{C4} & \textbf{C1} & \textbf{C2} & \textbf{C3} & \textbf{C4} & \textbf{C1} & \textbf{C2} & \textbf{C3} & \textbf{C4}  \\ 
			\toprule\toprule
			&                                                & \textcolor{blue}{\textbf{G*}}                                                                       & 15.36       & 21.96       & 24.55       & \textcolor{red}{\textbf{27.94}}           & 11.91       & 24.16       & 35.40       & 46.23       & 0.84        & 0.85        & 0.86        & 0.86        & 0.18        & 0.25        & 0.26        & 0.26         \\
			\cline{3-19}
			& \multirow{-2}{*}{Gemini}                                                & SG                                                                       & 15.77       & 22.55       & 26.55       & 27.54           & 12.20       & 23.83       & 35.47       & 44.02       & 0.84        & 0.85        & 0.86        & 0.86        & 0.18        & 0.24        & 0.26        & 0.27         \\ 
			\cline{2-19}
			&                                                    & \textcolor{blue}{\textbf{G*}}                                                                       & 26.40       & 43.31       & 51.29       & 59.08           & 8.26        & 15.95       & 23.48       & 29.50       & 0.92        & 0.92        & 0.92        & 0.93        & 0.74        & 0.76        & 0.77        & 0.78         \\
			\cline{3-19}
			& \multirow{-2}{*}{GMN}                                                   & SG                                                                       & 31.13       & 45.77       & 54.71       & \textcolor{red}{\textbf{59.68}}           & 3.78        & 15.67       & 22.8       & 28.13       & 0.92        & 0.92        & 0.92        & 0.93        & 0.75        & 0.77        & 0.78        & 0.79         \\  
			\cline{2-19}
			&                                  & G                                                                       & 34.33       & 48.70       & 54.49       & 56.89           & 10.35       & 16.54       & 20.49       & 23.85       & 0.88        & 0.91        & 0.91        & 0.92        & 0.46        & 0.52        & 0.53        & 0.53         \\ 
			\cline{3-19}
			\multirow{-6}{*}{\rotatebox[origin=c]{90}{\textsf{\textbf{Random}}}} &  \multirow{-2}{*}{SAFE} & SG & 37.13       & 51.89       & 58.08       & \textcolor{red}{\textbf{60.68}}           & 10.53       & 16.48       & 21.54       & 25.22       & 0.89        & 0.92        & 0.92        & 0.92        & 0.47        & 0.52        & 0.54        & 0.55         \\ 
			\hline\hline
			&                                                 & \textcolor{blue}{\textbf{G*}}                                                                       & 15.77       & 21.16       & 20.96       & 21.36           & 16.40       & 35.60       & 48.88       & 65.18       & 0.86        & 0.85        & 0.85        & 0.86        & 0.16        & 0.17        & 0.18        & 0.19         \\ 
			\cline{3-19}
			& \multirow{-2}{*}{Gemini}                                                & SG                                                                       & 15.17       & 18.56       & 19.36       & \textcolor{red}{\textbf{21.96}}           & 11.17       & 24.48       & 36.67       & 47.56       & 0.86        & 0.86        & 0.86        & 0.85        & 0.16        & 0.18        & 0.18        & 0.20         \\ 
			\cline{2-19}
			&                                                   & \textcolor{blue}{\textbf{G*}}                                                                       & 32.93       & 47.31       & 53.29       & \textcolor{red}{\textbf{53.29}}           & 6.82        & 16.50       & 22.81       & 26.68       & 0.92        & 0.92        & 0.93        & 0.94        & 0.59        & 0.65        & 0.68        & 0.70         \\
			\cline{3-19}
			& \multirow{-2}{*}{GMN}                                                   & SG                                                                       & 36.47       & 47.21       & 51.2       & 51.68           & 7.33        & 14.65       & 20.59       & 24.10       & 0.93        & 0.93        & 0.93        & 0.94        & 0.60        & 0.64        & 0.68        & 0.68         \\
			\cline{2-19}
			&                                  & G                                                                       & 60.28       & 77.05       & 80.84       & 83.43           & 20.36       & 32.90       & 47.90       & 55.36       & 0.90        & 0.91        & 0.92        & 0.92        & 0.38        & 0.43        & 0.45        & 0.46         \\ 
			\cline{3-19}
			\multirow{-6}{*}{\rotatebox[origin=c]{90}{\textsf{\textbf{Balanced}}}} & \multirow{-2}{*}{SAFE} & SG                      & 61.01       & 75.85       & 82.83       & \textcolor{red}{\textbf{83.43}}           & 9.49        & 14.24       & 18.02       & 21.28       & 0.90        & 0.92        & 0.93        & 0.93        & 0.39        & 0.44        & 0.46        & 0.46         \\
			\bottomrule
	\end{tabular}}
\end{table*}

\section{Evaluation}\label{sec:Eval}
In this section we evaluate our attacks and investigate the following research questions: 
\begin{mybox}
	\textbf{RQ1}: \textit{Are the three target models more robust against targeted or untargeted attacks?}

\textbf{RQ2}: \textit{Are the three target models more robust against black-box or white-box approaches?}

\textbf{RQ3}: \textit{Are models that consider CFG topology more robust against our attacks if compared to models that do not?}

\textbf{RQ4}: \textit{Does a different number of instructions for the source and target functions affect the success of the attack?}
\end{mybox}

\paragraph*{\textbf{Performance metrics}}
Our main evaluation metric is the \textbf{Attack success rate (A-rate)}, that is the percentage of adversarial samples that successfully mislead the target model.
We complement our investigation by collecting a set of support metrics to gain qualitative and quantitative insights into the attacking process: 
\begin{itemize}
\item \textbf{Modification size (M-size)}: number of inserted instructions;
\item \textbf{Average Similarity (A-sim)}: obtained final similarity values;
\item \textbf{Normalized Increment (N-inc)}: similarity increments normalized with respect to the initial value; only used for targeted attacks;
\item \textbf{Normalized Decrement (N-dec)}: similarity decrements normalized with respect to the initial value; only used for untargeted attacks.
\end{itemize}

Support metrics are computed over the set of samples that successfully mislead the model.

As an example, let us consider a targeted attack against three pairs of functions with initial similarities 0.40, 0.50, and 0.60. After the attack we reach final similarities that are 0.75, 0.88, and 0.94. We deem an attack as successful if the final similarity is above $0.8$ (the reason will be clear in the next section). In this example, we have an \textbf{A-rate} of 66.66\%, an \textbf{A-sim} of 0.91 and a \textbf{N-inc} of 0.81.

The \textbf{N-inc} is the average of the formula below over the samples that successfully mislead the model:

\begin{equation}\label{eq:N_inc}
\begin{aligned}
\frac{final \; similarity - initial \; similarity}{1 - initial \; similarity}
\end{aligned}
\end{equation}

The denominator for the fraction above is the maximum possible increment for the analyzed pair: we use it to normalize the obtained increment. Intuitively, the value of this metric is related with the initial similarities of the successfully attacked pair. Consider a targeted attack where a pair exhibits a final similarity of 0.80. When the normalized increment is 0.7, their initial similarity is 0.33 (from Equation~\ref{eq:N_inc}); when the normalized increment is 0.3, we have a much higher 0.7 initial similarity.

The comparison between  \textbf{A-sim} and the success threshold gives us insights on the ability of the attack to reach high similarity values. In the aforementioned example, the  \textbf{A-sim} value of 0.91 shows that when the attack is able to exceed the success threshold, it has actually an easy time to bring the similarity around the value of 0.91. 

\paragraph*{\textbf{Evaluation outline}}
We test our black-box and white-bock attacks against each target model in both scenarios. As discussed in Section~\ref{sec:dataset}, we use datasets \textsf{Random} and \textsf{Balanced} for the former and dataset \textsf{Untarg} for the latter.

\subsection{Setup}
In this section, we describe the attack parameters selected for our experimental evaluation.

\paragraph*{\textbf{Successful Attacks}} \label{sec:tau}

An attack is successful depending on the similarity value between the adversarial sample and the target function. For a targeted attack, the similarity score has to be increased during the attack until it trespasses a success threshold \T. For an untargeted attack, this score, which is initially 1, has to decrease until it is below a success threshold \U. Operatively, the values of such thresholds are determined by the way the similarity score is used in practice. In our experimental evaluation, we choose the thresholds as follows. We compute the similarity scores that our attacked systems give over a set of similar pairs and over a set of dissimilar pairs. For the first set, the average score is 0.79 with a standard deviation of 0.15. For the second set, these values are respectively 0.37 and 0.17. We thus opted for a success threshold $\mU=0.5$ for untargeted attacks and $\mT=0.8$ for targeted ones. Both \U and \T are over one standard deviation distant from the average similarity value measured for the relevant set for the attack. For the charts, we plot \U $\in[0.46, 0.62]$ and \T $\in[0.74, 0.88]$.

To fully understand the performance of the attacks, we also measure the amount of function pairs in a dataset already meeting a given threshold. For the targeted scenario, we plot it as a line labeled \textbf{C0}. As our readers can see (Figure~\ref{img:arates_BBTargeted} and~\ref{img:arates_WBTargeted}), their contribution is marginal: hence, we do not discuss them in the remainder of the evaluation. For the untargeted scenario, no such pair can exist by construction.

\paragraph*{\textbf{Black-box Attacks}}\label{sec:blacksetup}
To evaluate the effectiveness of Spatial Greedy against the black-box baseline Greedy, we select a maximum perturbation size $\bar{\delta}$ and a number of dead branches $B$ in four settings: \textbf{C1} ($\bar{\delta}=15$, $B=5$), \textbf{C2} ($\bar{\delta}=30$, $B=10$), \textbf{C3} ($\bar{\delta}=45$, $B=15$), and \textbf{C4} ($\bar{\delta}=60$, $B=20$).

We set $\varepsilon = 0.1$ in all greedy attacks. For Spatial Greedy and black-box Greedy, we test two sizes for the set of candidates: $110$ and $400$. For Greedy, we pick $110$ instructions manually and then randomly add others for a total of $400$; for Spatial Greedy, we recall that the selection is dynamic (Section~\ref{sec:spatialgreedy:selection}). The larger size brought consistently better results in both attacks, hence we present results only for it. Finally, for Spatial Greedy, we use $c = 10$ and $r \in \{0.25, 0.50, 0.75\}$, with $r = 0.75$ being the most effective choice in our tests (thus, the only one presented next). For the gray-box Greedy embodiments for Gemini and GMN, we refer to Section~\ref{sec:black:gemini} and~\ref{sec:black:gmn}, respectively.

\paragraph*{\textbf{White-box Attack}}\label{sec:whitesetup}
We evaluate GCAM considering four different values for the number $B$ of inserted dead branches: \textbf{C1} ($B=5$), \textbf{C2} ($B=10$), \textbf{C3} ($B=15$), and \textbf{C4} ($B=20$). For each model, we use the number of iterations that brings the attack to convergence. 

\begin{table*}[t!]
\ra{1.5}
\footnotesize
\centering
\caption{Evaluation metrics with $\mU=0.50$ relative to the \textbf{black-box} attacks against the three target models in the \textbf{untargeted} scenario. Spatial Greedy (SG) is evaluated using parameters $\varepsilon = 0.1$ and $r = 0.75$. Greedy (G) is evaluated using $\varepsilon = 0.1$. Similarly to Table~\ref{tab:BBTargeted}, \textcolor{blue}{\textbf{G*}} is the gray-box version of Greedy where applicable. When examining G against SAFE, a set of candidates of size 400 is considered.}
\label{tab:BBUntargeted}
\resizebox{\linewidth}{!}{%
	\begin{tabular}{c|c|c|cccc|cccc|cccc|cccc} 
		\toprule
		\multicolumn{1}{l|}{\multirow{2}{*}{}} & \multirow{2}{*}{\textbf{Target}} & \multirow{2}{*}{\textbf{Attack}}                        & \multicolumn{4}{c|}{\textbf{A-rate (\%) ($\mU=0.50$)}} & \multicolumn{4}{c|}{\textbf{M-size ($\mU=0.50$)}}  & \multicolumn{4}{c|}{\textbf{A-sim ($\mU=0.50$)}}   & \multicolumn{4}{c}{\textbf{N-dec ($\mU=0.50$)}}     \\ 
		\cline{4-19}
		\multicolumn{1}{l|}{} & \multicolumn{1}{l|}{} & \multicolumn{1}{l|}{} & \textbf{C1} & \textbf{C2} & \textbf{C3} & \textbf{C4}     & \textbf{C1} & \textbf{C2} & \textbf{C3} & \textbf{C4} & \textbf{C1} & \textbf{C2} & \textbf{C3} & \textbf{C4} & \textbf{C1} & \textbf{C2} & \textbf{C3} & \textbf{C4}  \\ 
		\toprule\toprule
		
		& & \textcolor{blue}{\textbf{G*}}                            & 23.60        & 37.40        & 47.0        & 50.60            & 2.73        & 5.24        & 9.16        & 11.78       & 0.47        & 0.48        & 0.48        & 0.48        & 0.53        & 0.52        & 0.52        & 0.52         \\ 
		\cline{3-19}
		& \multirow{-2}{*}{Gemini}      & SG                             & 22.95        & 40.32        & 48.10        & \textcolor{red}{\textbf{53.89}}            & 3.32        & 5.80        & 9.09        & 11.35       & 0.48        & 0.48        & 0.48        & 0.48        & 0.52        & 0.52        & 0.52        & 0.52         \\ 
		\cline{2-19}
		& & \textcolor{blue}{\textbf{G*}}                             & 68.40        & 86.20        & 91.02       & \textcolor{red}{\textbf{93.81}}           & 2.55        & 3.77        & 3.82        & 3.71        & 0.27        & 0.25        & 0.22        & 0.21        & 0.73        & 0.75        & 0.77        & 0.79         \\
		\cline{3-19}
		& \multirow{-2}{*}{GMN}         & SG                             & 65.87        & 83.23        & 88.13       & 91.62           & 2.75        & 3.21        & 4.12        & 4.14        & 0.27        & 0.24        & 0.24        & 0.23        & 0.73        & 0.76        & 0.76        & 0.77         \\
		\cline{2-19}
		& & G                             & 44.71       & 74.65       & 82.83       & 88.22           & 6.59        & 8.28        & 8.77        & 9.04        & 0.40        & 0.42        & 0.42        & 0.42        & 0.40        & 0.44        & 0.45        & 0.45         \\ 
		\cline{3-19}
		\multirow{-6}{*}{\rotatebox[origin=c]{90}{\textsf{\textbf{Untarg}}}} & \multirow{-2}{*}{SAFE} & SG & 56.49       & 80.83       & 87.42       & \textcolor{red}{\textbf{90.62}}           & 6.67        & 7.74        & 7.77        & 7.64        & 0.39        & 0.41        & 0.42        & 0.42        & 0.60        & 0.59        & 0.58        & 0.59         \\
		\bottomrule
\end{tabular}}
\end{table*}

\subsection{Complete Attack Results}\label{sec:complete-results}
This section provides complete results for our black-box and white-box attacks on the three target models. For brevity, we focus only on Spatial Greedy when discussing black-box targeted and untargeted attacks, leaving out the results for the baseline Greedy. The two will see a detailed comparison later, with Spatial Greedy emerging as generally superior.

\subsubsection{Black-box Targeted Attack}\label{sec:BB_Targeted}
Considering an attacker with {black-box} knowledge in a {targeted} scenario, the three target models show a similar behavior against Spatial Greedy.

\subsubsection*{\textsf{Random} Dataset}
\begin{figure}[t!]
\centering
\includegraphics[width=3.5in, trim = 0cm 0.6cm 0cm 0cm]{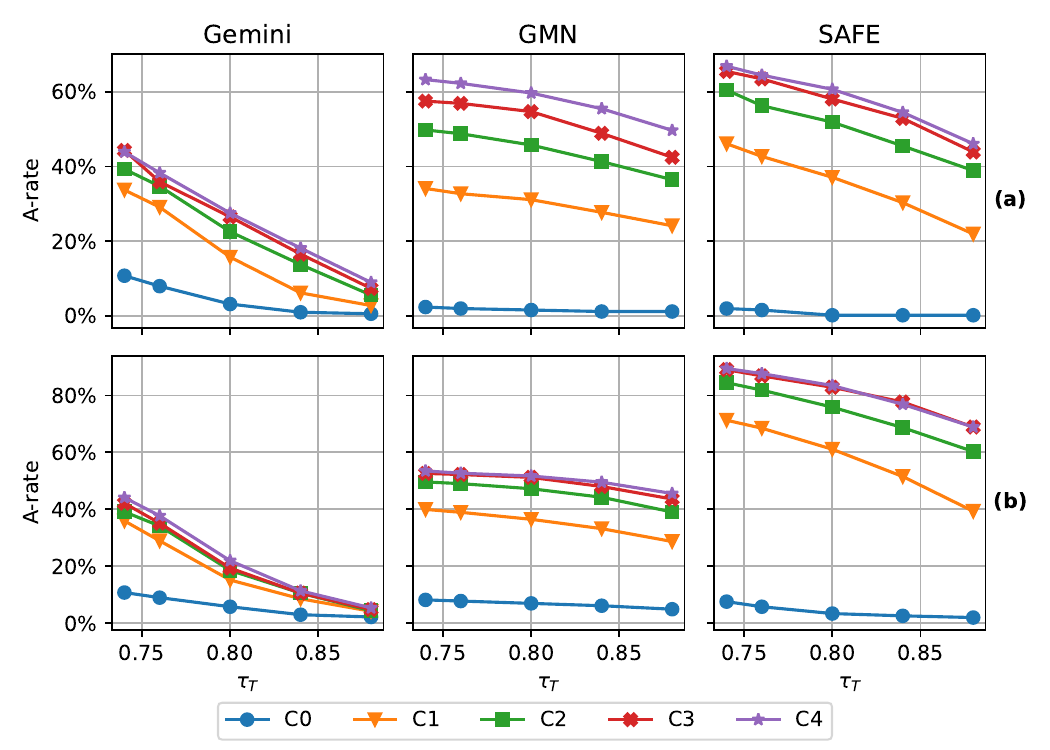}
\caption{\textbf{Black-box targeted} attack with Spatial Greedy against the three target models on the \textsf{Random} \textbf{(a)} and \textsf{Balanced} \textbf{(b)} datasets, while varying the success threshold \T $\in[0.74, 0.88]$. We use a set of candidates of 400 instructions, $\varepsilon = 0.1$, and $r = 0.75$.} 
\label{img:arates_BBTargeted}
\end{figure}

The attack success rate \textbf{A-rate} is positively correlated with the number $B$ of dead branches and the maximum number $\bar{\delta}$ of instructions introduced in the adversarial example. Fixing at $\mT=0.80$ the success threshold for the attack, we have an \textbf{A-rate} that on Gemini goes from 15.77\% (setting \textbf{C1}) up to 27.54\% (setting \textbf{C4}). The other target models follow this behavior, as the \textbf{A-rate} for GMN goes from 31.13\% up to 59.68\%, and from 37.13\% up to 60.68\% for SAFE. This trend holds for other success thresholds as visible in Figure~\ref{img:arates_BBTargeted}. From these results, it is evident that the higher the values of the two parameters, the lower the robustness of the attacked models. Table~\ref{tab:BBTargeted} presents a complete overview of the results.

The other metrics confirm the relationship between the parameters $B$ and $\bar{\delta}$ and the effectiveness of our attack. In particular, when increasing the perturbation size, as highlighted by the modification size \textbf{M-size} metric, both \textbf{A-sim} and the normalized increment \textbf{N-inc} increase, suggesting that incrementing the perturbation size is always beneficial. 

\subsubsection*{\textsf{Balanced Dataset}}
We see similar trends also when considering pairs of functions having similar initial length. When fixing $\mT =0.80$, the \textbf{A-rate} for GMN is as low as 36.47\% in setting  \textbf{C1} and as high as 51.68\% in setting \textbf{C4}; for SAFE, it is as low as 61.01\% in setting \textbf{C1} and as high as 83.43\% in setting \textbf{C4}. Contrarily, Gemini shows higher resilience to the attacks, as the \textbf{A-rate} referring to $\mT= 0.80$ is as low as 15.17\% in \textbf{C1} and as high as 21.96\% in setting \textbf{C4}. Figure~\ref{img:arates_BBTargeted} and Table~\ref{tab:BBTargeted} report a complete overview of the results.

The \textbf{A-sim} and the \textbf{N-inc} metrics further highlight the correlation between the two parameters and the efficacy of our approaches. When increasing the perturbation size, as pointed out by the \textbf{M-size} metric, it is easier to obtain higher final similarity values regardless of the initial ones.

\subsubsection{Black-box Untargeted Attack}\label{sec:BB_Untargeted}
Considering an attacker with {black-box} knowledge in a {untargeted} scenario, all the three target models are vulnerable to Spatial Greedy, with different robustness.

\begin{figure}[t!]
\centering
\includegraphics[width=3.5in, trim = 0cm 0.6cm 0cm 0cm]{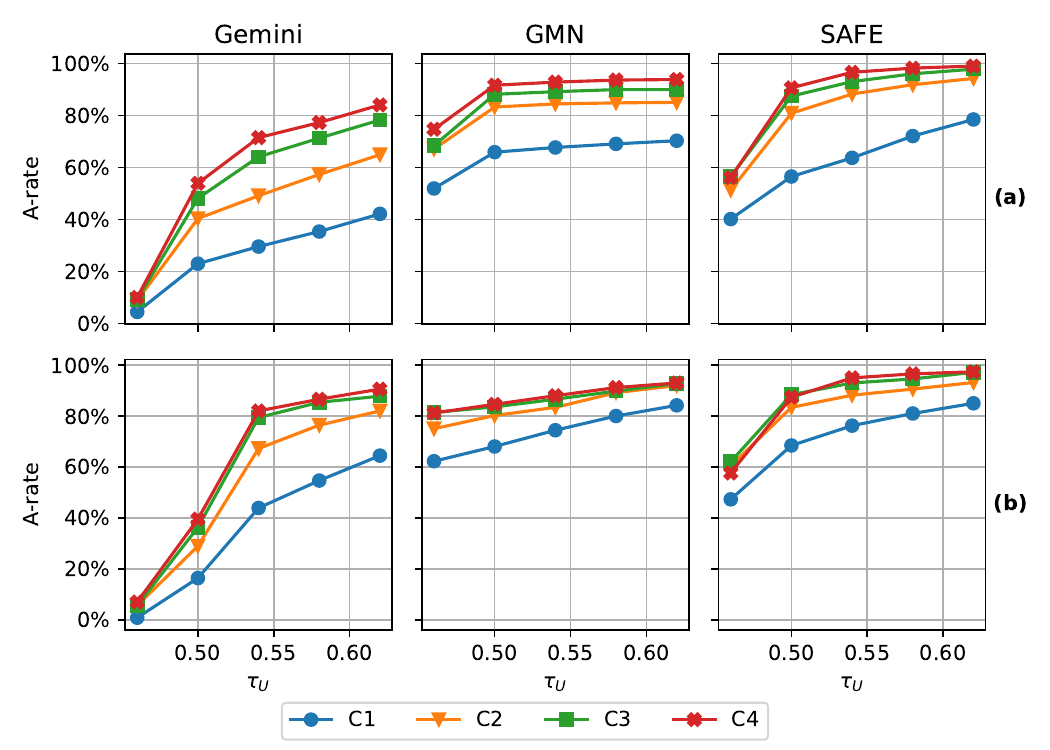}
\caption{\textbf{(a) Black-box untargeted} attack with Spatial Greedy against the three target models while varying the success threshold \U $\in[0.46, 0.62]$, and the settings \textbf{C1}, \textbf{C2}, \textbf{C3}, and \textbf{C4}. We use a set of candidates of 400 instructions, $\varepsilon = 0.1$, and $r = 0.75$. \newline \textbf{(b) White-box} \textbf{untargeted} attack against the three target models while varying the success threshold \U $\in[0.46, 0.62]$, and the settings \textbf{C1}, \textbf{C2}, \textbf{C3}, and \textbf{C4}. \textit{Left}: GCAM attack with 40k iterations against GEMINI. \textit{Center}: GCAM attack with 1k iterations against GMN. \textit{Right}: GCAM attack with 1k iterations against SAFE.}
\label{img:BB_WB_Untargeted}
\end{figure}

The observations highlighted in Section~\ref{sec:BB_Targeted} also hold in this scenario. Incrementing  $B$ and $\bar{\delta}$ is beneficial for the attacker. As visible in Figure~\ref{img:BB_WB_Untargeted} and in Table~\ref{tab:BBUntargeted}, the attack success rate \textbf{A-rate} for $\mU=0.50$ in setting \textbf{C1} is 22.95\% for Gemini, 65.87\% for GMN, and 56.49\% for SAFE. The metric increases across settings, peaking at 53.89\% for Gemini, 91.62\% for GMN, and 90.62\% for SAFE in setting \textbf{C4}.

Table~\ref{tab:BBUntargeted} also reports the results for modification size metric \textbf{M-size}. In this case, we can see the effectiveness of Spatial Greedy as a small number of added instructions is needed against each of the considered target models. Indeed, considering setting \textbf{C4}, which is the one that modifies the function most, the \textbf{M-size} at $\mU=0.50$ is 11.35 for Gemini, 4.14 for GMN, and 7.64 for SAFE.

\subsubsection{White-box Targeted Attack}\label{sec:WB_Targeted}

With an attacker with {white-box} knowledge in a {targeted} scenario, the three target models show different behaviors. Table~\ref{tab:WBTargeted} presents a complete overview of the results.

\begin{table*}
\centering
\footnotesize
\ra{1.7}
\caption{Evaluation metrics with $\mT=0.80$ for the \textbf{white-box targeted} attack against the three target models. The GCAM attack is executed up to 20k iterations for Gemini and up to 1k for GMN and SAFE.}
\label{tab:WBTargeted}
\resizebox{\linewidth}{!}{%
	\begin{tabular}{c|c|cccc|cccc|cccc|cccc} 
		\toprule
		\multicolumn{1}{l|}{\multirow{2}{*}{}} & \multicolumn{1}{l|}{\multirow{2}{*}{\textbf{Target}}} & \multicolumn{4}{c|}{\textbf{A-rate (\%) ($\mT=0.80$)}}                                                        & \multicolumn{4}{c|}{\textbf{M-size ($\mT=0.80$)}}                       & \multicolumn{4}{c|}{\textbf{A-sim ($\mT=0.80$)}}                                                            & \multicolumn{4}{c}{\textbf{N-inc ($\mT=0.80$)}}                                           \\ 
		\cline{3-18}
		\multicolumn{1}{l|}{}                  & \multicolumn{1}{l|}{}                                 & \textbf{C1}                    & \textbf{C2}                     & \textbf{C3} & \textbf{C4}                     & \textbf{C1}                      & \textbf{C2} & \textbf{C3} & \textbf{C4} & \textbf{C1}                    & \textbf{C2}                    & \textbf{C3} & \textbf{C4}                    & \textbf{C1} & \textbf{C2} & \textbf{C3}                    & \textbf{C4}                     \\ 
		\toprule\toprule
		& Gemini                                                & 24.35                          & 29.94                           & 30.94       & \textcolor{red}{\textbf{31.60}}        & 53.67  & 86.53       & 111.83      & 133.84      & 0.85                           & 0.86                           & 0.86        & 0.86      & 0.54        & 0.60        & 0.63       & 0.62                            \\ 
		\cline{2-18}
		& GMN                                                   & 34.73            & \textcolor{red}{\textbf{38.32}}             & 34.93       & 35.33      & 212.79      & 350.5       & 439.03      & 461.08      & 0.85                           & 0.84                           & 0.84        & 0.84         & 0.78        & 0.78        & 0.77          & 0.79  \\
		\cline{2-18}
		\multirow{-3}{*}{\rotatebox[origin=c]{90}{\textsf{\textbf{Random}}}}    & SAFE                           & 11.57                          & 18.96                           & 20.36       & \textcolor{red}{\textbf{21.76}}                           & 18.62                            & 31.87       & 38.27       & 38.90       & 0.84                           & 0.85                           & 0.85        & 0.85         & 0.67        & 0.71        & 0.70                           & 0.71  \\ 
		\hline\hline
		& Gemini                                 & \textcolor{red}{\textbf{36.60}}        & 34.93     & 33.13       & 29.94      & 95.86            & 101.83      & 134.51      & 161.63      & 0.86       & 0.85          & 0.86        & 0.86           & 0.53        & 0.51        & 0.52       & 0.54                            \\ 
		\cline{2-18}
		& GMN                                                   & 44.71                          & \textcolor{red}{\textbf{55.89}}    & 49.30       & 45.31        & 474.73            & 688.89      & 838.48      & 847.01      & 0.85       & 0.85   & 0.85        & 0.85      & 0.70        & 0.72        & 0.71        & 0.71                            \\
		\cline{2-18}
		\multirow{-3}{*}{\rotatebox[origin=c]{90}{\textsf{\textbf{Balanced}}}}                & SAFE                                                  & 16.60                           & 26.34                           & 25.35       & \textcolor{red}{\textbf{27.35}} & 16.39  & 28.30       & 33.67       & 36.99       & 0.85                           & 0.84                           & 0.85        & 0.85      & 0.47     & 0.56        & 0.57                           & 0.58                            \\
		\bottomrule
\end{tabular}}
\end{table*}

\begin{figure}[t!]
\centering
\includegraphics[width=3.5in, trim = 0cm 0.6cm 0cm 0cm]{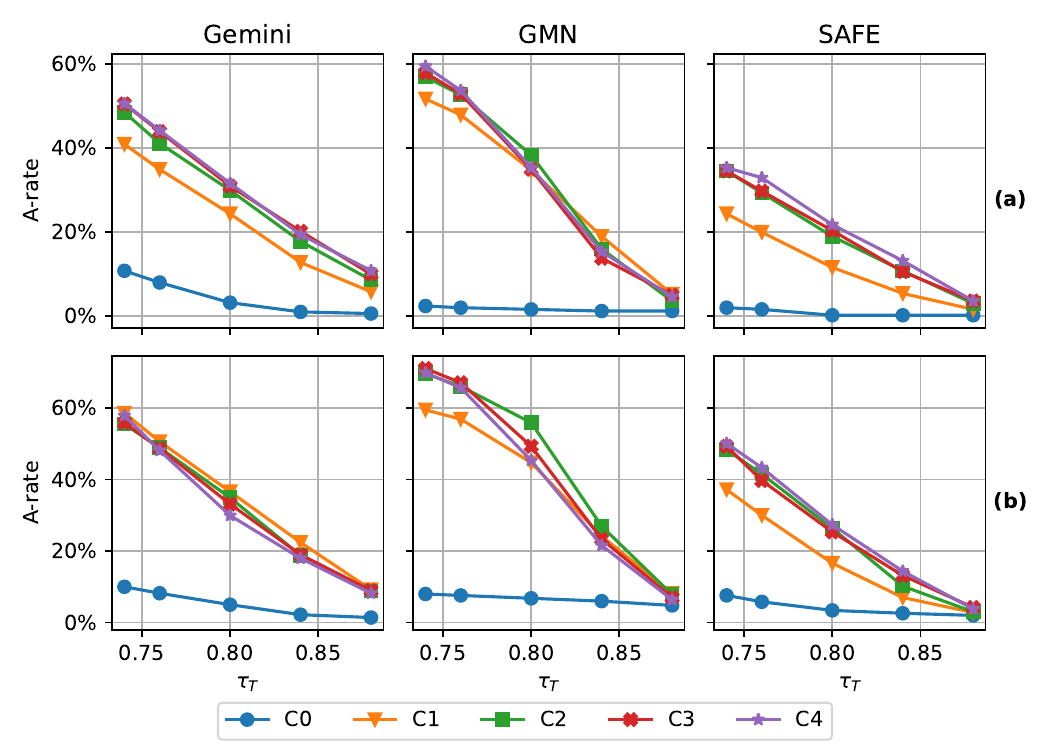}
\caption{\textbf{White-box} \textbf{targeted} attack against the three target models on the \textsf{Random} \textbf{(a)} and \textsf{Balanced} \textbf{(b)} datasets while varying the success threshold \T $\in[0.74, 0.88]$ and settings \textbf{C0} to \textbf{C4}. \textit{Left}: GCAM attack with 20k iterations against GEMINI. \textit{Center}: GCAM attack with 1k iterations against GMN. \textit{Right}: GCAM attack with 1k iterations against SAFE.} 
\label{img:arates_WBTargeted}
\end{figure}

\subsubsection*{\textsf{Random Dataset}}
Both Gemini and SAFE show a higher robustness to our GCAM attack if compared to GMN.

As visible in Figure~\ref{img:arates_WBTargeted}, when attacking Gemini and SAFE, there is a positive correlation between the number $B$ of locations (i.e., dead branches) where to insert perturbations and the attack success rate \textbf{A-rate}. When considering setting \textbf{C1}, the \textbf{A-rate} for $\mT=0.80$ is 24.35\% for Gemini, and 11.57\% for SAFE; moving to \textbf{C4}, it increases up to 31.60\% for Gemini, and 21.76\% for SAFE. On the contrary, GMN does not show a monotonic \textbf{A-rate} increase for an increasing $B$ value, as the peak \textbf{A-rate} is 38.32\% in setting \textbf{C2}.

We now discuss the modification size \textbf{M-size} metric: fixing $\mT=0.80$ and considering the setting where \textbf{A-rate} peaks, we measure an \textbf{M-size} value of 38.90 for SAFE (\textbf{C4}), 133.84 for Gemini (\textbf{C4}), and 350.50 for GMN (\textbf{C2}): SAFE is the model that sees the insertion of fewer instructions. This is not surprising: due to the feature-space representation of SAFE,  the embeddings we alter in the attack for it (Section~\ref{sec:white:safe}) refer to a number of instructions that is fixed.

\begin{table*}[t!]
\centering
\footnotesize
\ra{1.5}
\caption{Evaluation metrics with $\mU=0.50$ for the \textbf{white-box untargeted} attack against the three target models. The GCAM attack is executed up to 20k iterations for Gemini and up to 1k for GMN and SAFE.}
\label{tab:WBUntargeted}
\resizebox{\linewidth}{!}{%
	\begin{tabular}{c|c|cccc|cccc|cccc|cccc} 
		\toprule
		\multicolumn{1}{l|}{\multirow{2}{*}{}} & \multicolumn{1}{l|}{\multirow{2}{*}{\textbf{Target}}} & \multicolumn{4}{c|}{\textbf{A-rate (\%) ($\mT=0.50$)}}                                                        & \multicolumn{4}{c|}{\textbf{M-size ($\mT=0.50$)}}                       & \multicolumn{4}{c|}{\textbf{A-sim ($\mT=0.50$)}}                                                            & \multicolumn{4}{c}{\textbf{N-dec ($\mT=0.50$)}}                                           \\ 
		\cline{3-18}
		\multicolumn{1}{l|}{}                  & \multicolumn{1}{l|}{}                                 & \textbf{C1}                    & \textbf{C2}                     & \textbf{C3} & \textbf{C4}                     & \textbf{C1}                      & \textbf{C2} & \textbf{C3} & \textbf{C4} & \textbf{C1}                    & \textbf{C2}                    & \textbf{C3} & \textbf{C4}                    & \textbf{C1} & \textbf{C2} & \textbf{C3}                    & \textbf{C4}                     \\ 
		\toprule\toprule
		
		& Gemini                                                & 16.37       & 28.94       & 36.32                           & \textcolor{red}{\textbf{39.52}}       & 46.63       & 79.76       & 103.86          & 117.21      & 0.48        & 0.48        & 0.48        & 0.47        & 0.51        & 0.52        & 0.52        & 0.53         \\ 
		\cline{2-18}
		& GMN                                                   & 68.06       & 80.24       & 83.63                           & \textcolor{red}{\textbf{84.63}}       & 298.34      & 541.32      & 718.57       & 859.53      & 0.23        & 0.19        & 0.18        & 0.18        & 0.78        & 0.81        & 0.82        & 0.82         \\
		\cline{2-18}
		\multirow{-3}{*}{\rotatebox[origin=c]{90}{\textsf{\textbf{Untarg}}}} & SAFE                                                  & 68.46       & 83.43       & \textcolor{red}{\textbf{88.42}}      &  87.42       & 17.52       & 30.01       & 36.0       & 38.49       & 0.39        & 0.38        & 0.39        & 0.40        & 0.56        & 0.58        & 0.58        & 0.58         \\
		\bottomrule
\end{tabular}}
\end{table*}

\subsubsection*{\textsf{Balanced Dataset}}
On pairs of functions having similar initial length, the GCAM attack can generate adversarial examples that can deceive all the three target models, even though we find no direct relationship between the parameter $B$ and the attack success rate \textbf{A-rate}. As visible in Figure~\ref{img:arates_WBTargeted} and reported in Table~\ref{tab:WBTargeted}, SAFE manifests the highest robustness with a peak \textbf{A-rate} of 27.35\% in setting \textbf{C4} at $\mT=0.80$. For the same threshold choice, the peak \textbf{A-rate} is higher for both Gemini (i.e., 36.60\% in \textbf{C1}) and GMN (i.e., 55.89\% in \textbf{C2}).

Consistently with the \textsf{Random} results, GCAM resorts to fewer instructions when targeting the SAFE model: the modification size \textbf{M-size} at $\mT=0.80$ is 36.99 in the \textbf{C4} setting. On the contrary, the \textbf{M-size} at $\mT=0.80$ is 95.86 in setting \textbf{C1} for Gemini, and 688.89 in setting \textbf{C2} for GMN.

\subsubsection{White-box untargeted attack}\label{sec:WB_Untargeted}
Figure~\ref{img:BB_WB_Untargeted} and Table~\ref{tab:WBUntargeted} report the results for our attacks with {white-box} knowledge in the {untargeted} scenario.

Gemini looks more robust than the other models: for example, fixing $\mU=0.50$, we measure the highest attack success rate \textbf{A-rate} as 39.52\% in the \textbf{C4} setting. On the contrary, for the same \U, the highest \textbf{A-rate} for SAFE is 88.42\% (setting \textbf{C3}) and 84.63\% for GMN (setting \textbf{C4}).

The general trend of having a positive correlation of $B$ and the \textbf{A-rate} is still observable (with a sharp increase of the \textbf{A-rate} from setting \textbf{C1} to \textbf{C2}). The \textbf{M-size} shows that SAFE is the most fragile model in terms of instructions to add, as they are much fewer than with the other two models. 

\subsubsection{Greedy vs. Spatial Greedy}
\begin{figure}[h!]
\centering
\includegraphics[width=3.5in, trim = 0cm 0.6cm 0cm 0cm]{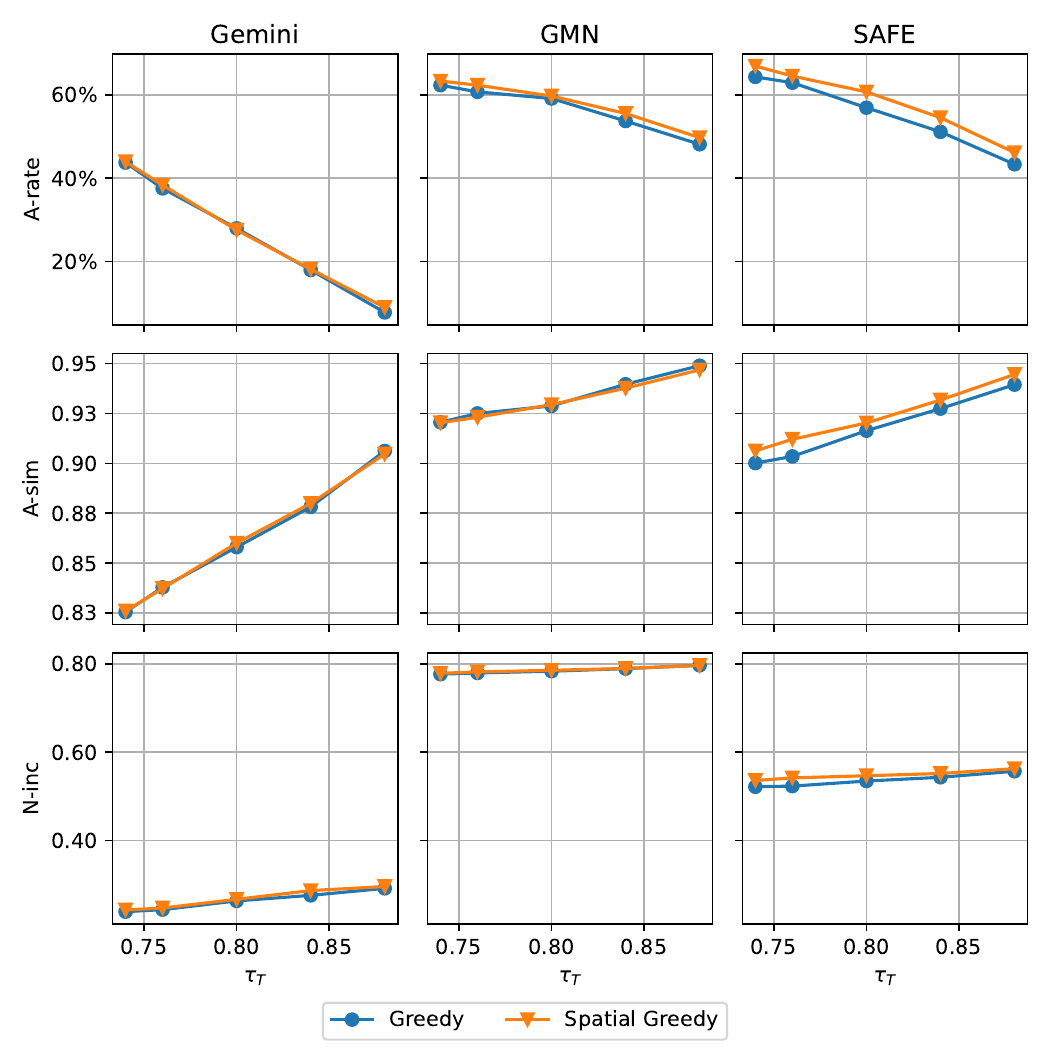}
\caption{\textbf{Greedy} and \textbf{Spatial Greedy} targeted attacks against the three models while varying the success threshold \T $\in[0.74, 0.88]$, using the \textsf{Random} dataset and setting \textbf{C4}. For both, we consider $\varepsilon = 0.1$ and $|\texttt{CAND}| = 400$. For Spatial Greedy, we also set $r = 0.75$.}
\label{img:greedySpatial_UB_BBTargeted}
\end{figure}

We now compare the performance of Spatial Greedy against the Greedy baseline, until now left out of our discussions for brevity. Figure~\ref{img:greedySpatial_UB_BBTargeted} shows the results for a targeted attack on the \textsf{Random}. Additional data points are available in Table~\ref{tab:BBTargeted}.

We discuss Gemini and GMN first. We recall that we could exploit their feature extraction process to reduce the size of the set of candidates, devising a gray-box Greedy procedure. Spatial Greedy is instead always black-box.

Considering the attack success rate \textbf{A-rate} at $\mT=0.80$, Spatial Greedy always outperform the gray-box baseline, except for setting \textbf{C4} on Gemini (although the two perform similarly: 27.94\% for Greedy and 27.56\% for Spatial Greedy).  Looking at the other metrics, we can see that our black-box approach based on instructions embeddings is on par or improves on the results provided by the gray-box baseline. There are only few cases in which gray-box Greedy outperforms Spatial Greedy: i.e., the modification size \textbf{M-size} at $\mT=0.80$ on Gemini for settings \textbf{C1} and \textbf{C3}.

Moving to SAFE, we recall that only a black-box Greedy is feasible. Considering the \textbf{A-rate}, we can notice that increasing both $\bar{\delta}$ and $B$ produces a more noticeable difference between the baseline technique and Spatial Greedy. In the \textbf{C1} setting, the \textbf{A-rate} at $\mT=0.80$ is 34.33\% for Greedy and 37.13\% for Spatial Greedy; then, it increases up to 56.89\% for Greedy and 60.68\% for Spatial Greedy when considering the \textbf{C4} scenario.

The other metrics confirm this behavior. Considering the average similarity \textbf{A-sim}, regardless of the chosen $\bar{\delta}$ and $B$ from the setting, we can observe that adversarial pairs generated through Spatial Greedy present a final average similarity that is higher than the one relative to the pairs generated using the baseline solution. The effectiveness of Spatial Greedy is finally confirmed by the normalized increment \textbf{N-inc} metric; at a comparison of the results, the impact of the candidates selected using Spatial Greedy is more consistent if compared to the one of the candidates selected using the baseline approach.

Comparing Spatial Greedy with Greedy, we measure on the \textsf{Random} dataset an average A-rate increase of $2.26$ and a decreased M-size by $0.5$ instructions across all configurations and models. On the \textsf{Balanced} dataset, the two attacks are essentially on par in terms of success ($0.25$ A-rate increase for Spatial Greedy), but Spatial Greedy uses appreciably fewer instructions (M-size decrease of $12$). 

We omit a detailed discussion for the untargeted scenario for brevity. For average results across all models and configurations, Spatial Greedy sees an A-rate increase of $1.75$, whereas the M-size is smaller by $0.16$ instructions.

\begin{mybox2}{\bf Take away:}
We conclude that \textbf{Spatial Greedy} is typically superior (and always at least comparable) to a Greedy attack even when an efficient gray-box Greedy variant is possible. The results suggest that our dynamic update of the set of candidates, done at each iteration of the optimization procedure, can lead to the identification of new portions of the instruction space (and consequently a new subset of the ISA) that can positively influence the attack results.
\end{mybox2}

\subsection{RQ1: Targeted vs. Untargeted Attacks}
From the previous sections, the attentive reader may have noticed that all our approaches are much more effective in an \textbf{untargeted} scenario for all models and proposed metrics. The analysis provided below takes, for the targeted attacks, the values obtained on the \textsf{Random} dataset, but analogous trends can be observed if picking the \textsf{Balanced} one instead.

When looking at attack success rate \textbf{A-rate} for all thresholds of similarities, the three target models are less robust against untargeted attacks (rather than targeted ones) regardless of the adversarial knowledge. For the best attack among black-box and white-box configurations, in the targeted scenario, the peak \textbf{A-rate} at $\mT=0.80$ is 27.54\% for Gemini, 59.68\% for GMN, and 60.68\% for SAFE. For the untargeted scenario, the peak \textbf{A-rate} at $\mU=0.50$ is 53.89\% for Gemini, 91.62\% for GMN, and 90.62\% for SAFE.

The number of instructions \textbf{M-size} needed for generating valid adversarial examples further confirms the weak resilience of the target models to untargeted attacks. When considering the worst setting according to \textbf{M-size} (i.e., \textbf{C4}), while we need only few instructions for untargeted attacks at $\mU=0.50$ (i.e., 11.35 for Gemini, 4.14 for GMN, and 7.64 for SAFE), we need a significantly higher number of added instructions for targeted attacks (i.e., 44.02 for Gemini, 28.13 for GMN, and 25.22 for SAFE) at $\mT=0.80$.

\begin{mybox2}{\bf Take away:}
On all the attacked models, both targeted and untargeted attacks are feasible, especially using Spatial Greedy (see also RQ2). Their resilience against untargeted attacks is significantly lower.
\end{mybox2} 

\begin{figure}[t!]
\centering
\includegraphics[width=3.5in, trim = 0cm 0.6cm 0cm 0cm]{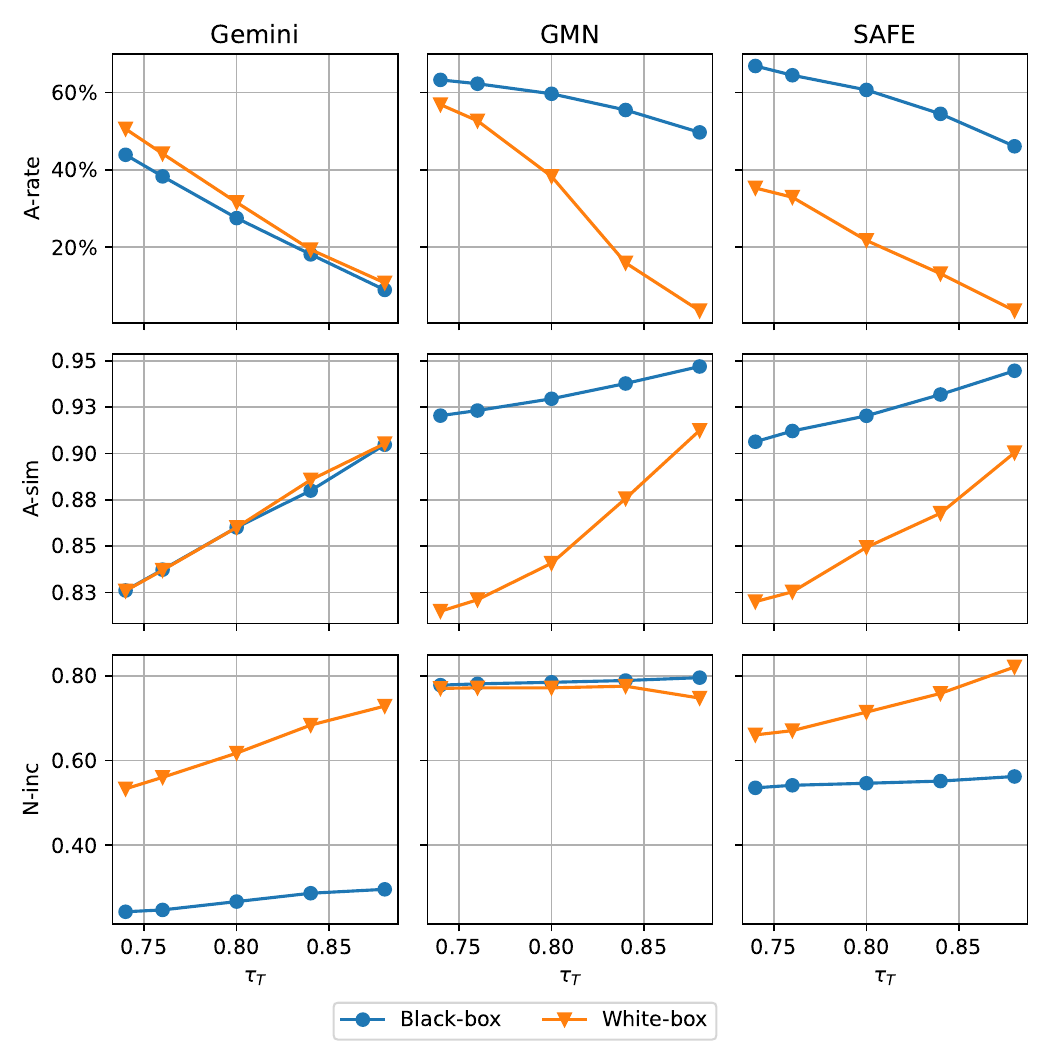}
\caption{\textbf{Black-box} and \textbf{white-box targeted} attacks against the three models while varying the success threshold \T $\in[0.74, 0.88]$ and considering the \textsf{Random} dataset. In the {black-box} scenario, all the results refer to the Spatial Greedy approach ($\varepsilon = 0.1$, $r = 0.75$, and $|\texttt{CAND}|=400$). In the {white-box} scenario, the results for Gemini are for a GCAM attack with 20k iterations while the ones for SAFE and GMN are for a GCAM attack with 1k iterations. We consider all approaches in their most effective parameter choice, being it always setting \textbf{C4} except for the GCAM attack against GMN, for which we consider setting \textbf{C2}.} 
\label{img:blackWhite_Targeted}
\end{figure}

\subsection{RQ2: Black-box vs. White-box Attacks}\label{ssec:rq2}
An interesting finding from our tests is that the white-box strategy does not always outperform the black-box one.

Figure~\ref{img:blackWhite_Targeted} depicts a comparison in the targeted scenario between Spatial Greedy and GCAM for the attack success rate \textbf{A-rate}, average similarity \textbf{A-sim}, and normalized increment \textbf{N-inc} metrics. The comparison considers the \textsf{Random} dataset; we will cover the \textsf{Balanced} dataset separately with a brief discussion. The figure shows how different values of the success attack threshold \T can influence the considered metrics. On GMN and SAFE, Spatial Greedy is more effective than GCAM, resulting in significantly higher \text{A-rate} values, while the two perform similarly on Gemini.

Interestingly, in contrast with the evaluation based on the \textbf{A-rate} metric, both the \textbf{A-sim} and \textbf{N-inc} values highlight a coherent behavior among the three target models. Generally, adversarial examples generated using Spatial Greedy exhibit a higher \textbf{A-sim} value than the white-box ones (considering $\mT=0.80$, we have 0.86 vs. 0.86 for Gemini, 0.93 vs. 0.84 for GMN, and 0.92 vs. 0.85 for SAFE). Looking at \textbf{N-inc}, we face a completely reversed situation; the metric is better in the adversarial samples generated using GCAM (0.62 for Gemini, 0.79 for GMN, and 0.71 for SAFE) compared to those from Spatial Greedy (0.27 for Gemini, 0.79 for GMN, and 0.55 for SAFE). These two observations lead us to the hypothesis that the black-box attack is more effective against pairs of binary functions that exhibit high initial similarity values and can potentially reach a high final similarity. On the other side, GCAM is particularly effective against pairs that are very dissimilar at the beginning.

For the \textsf{Balanced} dataset, we observe trends analogous to those discussed above for the effects of black-box and white-box targeted attacks. The only difference worth mentioning involves the \textbf{A-rate} for GMN, with the GCAM attack now prevailing on Spatial Greedy (55.89\% vs 51.68\%).

For the untargeted scenario, our results (Tables~\ref{tab:BBUntargeted} and~\ref{tab:WBUntargeted}) for the \textbf{A-rate} metric considering $\mU=0.50$ show that Spatial Greedy has a slight advantage on GCAM. For Spatial Greedy, we have best-setting values of 53.89\% for Gemini, 91.62\% for GMN, and 90.62\% for SAFE; for GCAM, we have 39.52\% for Gemini, 84.63\% for GMN, and 88.42\% for SAFE.

In our experiments, GCAM performed worse than the black-box strategy, which may look puzzling since theoretically a white-box attack should be more potent than a black-box one. We explored if this could be due to the obfuscated gradient phenomenon~\cite{athalye2018obfuscated} or the inverse feature mapping problem (Section~\ref{sec:invMapping}). Our hypothesis leans towards the latter.
To this end, we conducted a GCAM attack exclusively in the feature space by eliminating all constraints needed to identify a valid potential sample in the problem space (i.e., non-negativity of coefficients for Gemini and GMN, rounding to genuine instruction embeddings for SAFE). As a result, GCAM achieved a success rate between 86.03\% and 99.81\% in targeted scenarios and between 97.01\% and 100\% in untargeted ones.
As indicated in~\cite{athalye2018obfuscated}, high performance once constraints are removed strongly suggests the absence of obfuscated gradient phenomena.

\begin{mybox2}
{\bf Take away:} Our tests show that the Spatial Greedy black-box strategy is on par or beats our white-box GCAM attack based on a rounding inverse strategy. Further investigation is needed to confirm if this result will hold for more refined inverse feature mapping techniques and when attacking other models. 
\end{mybox2}

\subsection{RQ3: Role of CFG-related Features}\label{sec:rq-cfg}
Among the target models under study, we can distinguish between the ones considering the CFG during the feature extraction process (Gemini and GMN) and the one that does not (SAFE). We want to investigate whether this aspect can influence the robustness of a model against our attacks. We focus on the \textsf{Random} dataset as its pairs of functions have an average CFG node count difference that is higher than in the \textsf{Balanced} dataset (17.8 vs 7.5, Section~\ref{sec:dataset}). This metric acts as a proxy for having (more) different CFG-related features.

In the {targeted} black-box scenario (Figure~\ref{img:arates_BBTargeted} and Table~\ref{tab:BBTargeted}), SAFE is the weakest among the three considered models, as the peak attack success rate \textbf{A-rate} at $\mT=0.80$ is 60.68\% for SAFE, 59.68\% for GMN, and 27.54\% for Gemini (\textbf{C4} setting). We thus conducted a manual analysis of the CFGs of each tested pair of functions: the majority of the successful attacks on all three models are from pairs where the initial CFGs have a similar number of nodes.

In the targeted {white-box} scenario (Figure~\ref{img:arates_WBTargeted} and Table~\ref{tab:WBTargeted}), the results slightly change. In particular, GCAM is particularly effective against GMN (the \textbf{A-rate} for it at $\mT=0.80$ peaks at 38.32\%) while both SAFE and Gemini show more robustness (the peak \textbf{A-rate} $\mT=0.80$ is 21.75\% for SAFE and 31.60\% for Gemini). We believe that these results are due to the difficulty in inverting the feature mapping which, as pointed out in Section~\ref{sec:invMapping}, prevents the application of classical white-box approaches against code models.

As for untargeted attacks, we recall the initial CFGs in each pair are trivially identical. We find that the choice whether to rely on CFG features in the model does not impact the final result. In the black-box scenario, for example, in the peak setting \textbf{C4} the \textbf{A-rate} at $\mU=0.50$ is 53.89\% for Gemini, 91.62\% for GMN, and 90.62\% for SAFE (Table~\ref{tab:BBUntargeted} and Figure~\ref{img:BB_WB_Untargeted}). Similarly, in the white-box scenario, the \textbf{A-rate} at $\mU=0.50$ in the best setting is 39.52\% for Gemini, 84.63\% for GMN, and 88.42\% for SAFE (Table~\ref{tab:WBUntargeted} and Figure~\ref{img:BB_WB_Untargeted}).

\begin{mybox2}{\bf Take away:} The presence of CFG-based features in the attacked model plays an important role when manipulating pairs of functions that are unbalanced in the number of CFG nodes, requiring a greater effort by the attacker. Untargeted attacks are unaffected. \end {mybox2}

\begin{figure}[t!]
\centering
\includegraphics[width=3.5in, trim = 0cm 0.6cm 0cm 0cm]{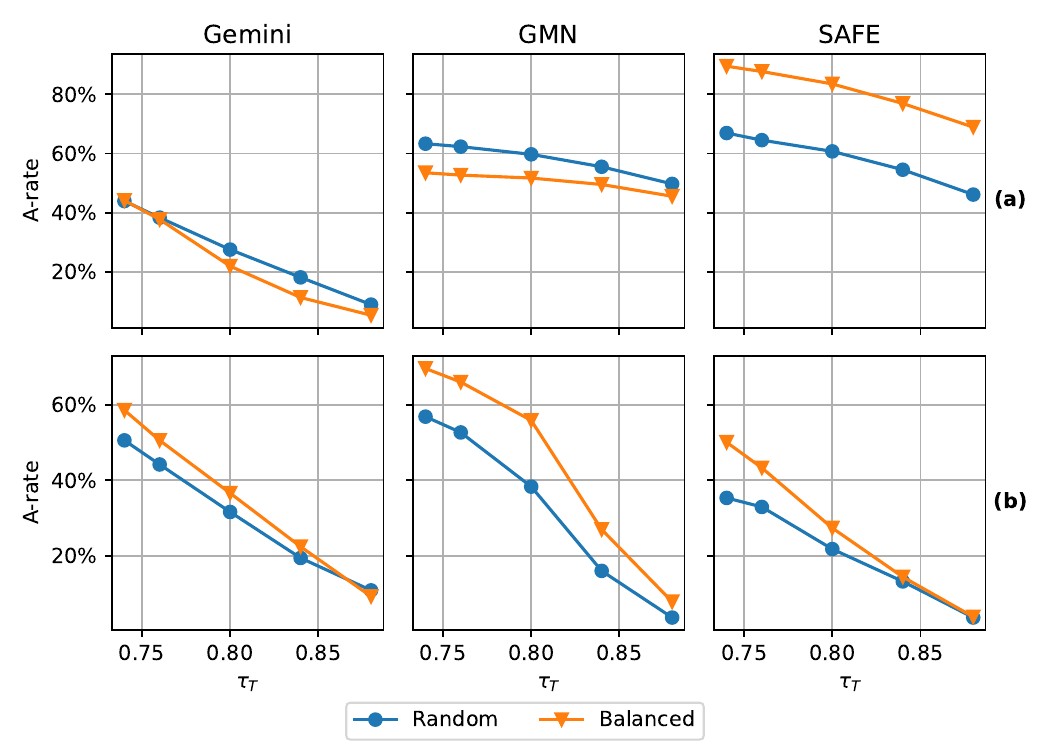}
\caption{Resilience of the three models to \textbf{targeted} \textbf{black-box} and \textbf{white-box} attackers on the \textsf{Random} and \textsf{Balanced} datasets for a success threshold \T $\in[0.74, 0.88]$. The best setting for each attack is shown.\newline 
\textbf{(a) Black-box} attacker. We test the Spatial Greedy approach against the target models with $\varepsilon = 0.1$, $r = 0.75$, $|\texttt{CAND}| = 400$, and setting \textbf{C4}.\newline
\textbf{(b) White-box} attacker. \textit{Left}: GCAM attack with 20k iterations against GEMINI in setting \textbf{C4} for the \textsf{Random} dataset and \textbf{C1} for the \textsf{Balanced} dataset. \textit{Center}: GCAM attack with 1k iterations against GMN, considering setting \textbf{C2} for both datasets. \textit{Right}: GCAM attack with 1k iterations against SAFE, considering setting \textbf{C4} for both datasets.} 
\label{img:balUnbal}
\end{figure}

\subsection{RQ4: Role of Initial Function Size Difference}\label{sec:rq-size}
In this section, we review our targeted-attack results by studying if the three target models are more robust against adversarial samples generated from pairs of either random functions (\textsf{Random} dataset) or functions initially balanced in length ({\textsf{Balanced} dataset}). We evaluate this research question by considering the attack success rate \textbf{A-rate}.

Figure~\ref{img:balUnbal} and Table~\ref{tab:BBTargeted} show the results in the case of a {black-box} attacker. SAFE is less resilient against adversarial samples generated from the \textsf{Balanced} dataset. For example, considering the \textbf{C4} setting, the \textbf{A-rate} at $\mT=0.80$ is 60.68\% on the \textsf{Random} dataset and it increases to 83.43\% on the \textsf{Balanced} dataset.  The other two models exhibit different behaviors: in the same settings as above, the \textbf{A-rate} is 27.54\% vs 21.96\% for Gemini and 59.68\% vs 51.68\% for GMN for pairs from \textsf{Random} and \textsf{Balanced}, respectively.

Figure~\ref{img:balUnbal} and Table~\ref{tab:WBTargeted} show the results in the case of {white-box} attacker. Interestingly, the three target models behave similarly in this scenario, showing better robustness against adversarial samples crafted from the \textsf{Random} dataset rather than from the \textsf{Balanced} one. Actually, the \textbf{A-rate} at $\mT=0.80$ is 31.60\% vs 36.6\% for Gemini, 38.32\% vs 55.89\% for GMN, and 21.76\% vs 27.35\% for SAFE.

\begin{mybox2}{\bf Take away:} Mounting our targeted attacks using a source function similar in length to the target one may or not be helpful depending on the model. The resilience of SAFE considerably drops in our tests, showing that an adversary has an easy time to create an adversarial sample. Conversely, the attacks we mount on pairs of functions with similar initial length are not more effective against Gemini and GMN than those on pairs of functions unbalanced in length.
\end{mybox2}

\section{Mirai Case Study}\label{sec:rq-mirai}
We complement our evaluation with a case study examining our attacks in the context of disguising functions from malware, analogously to exemplary scenario (1) from Section~\ref{sec:introduction}. 

We consider the code base from a famous leak of the Mirai malware, compiling it \texttt{gcc} 9.2.1 with  \texttt{-O0} optimization level on Ubuntu 20.04. After filtering out all functions with less than six instructions, we obtain a set of 63 functions. We build distinct datasets for the targeted and untargeted case. For the former, we pair malicious Mirai functions with benign ones from the \textsf{Random} dataset from the main evaluation, without making any attempt to balance lengths. For the latter, each of the 63 functions is paired with itself.

\begin{figure}[h!]
\centering
\includegraphics[width=3.5in, trim = 0cm 0.6cm 0cm 0cm]{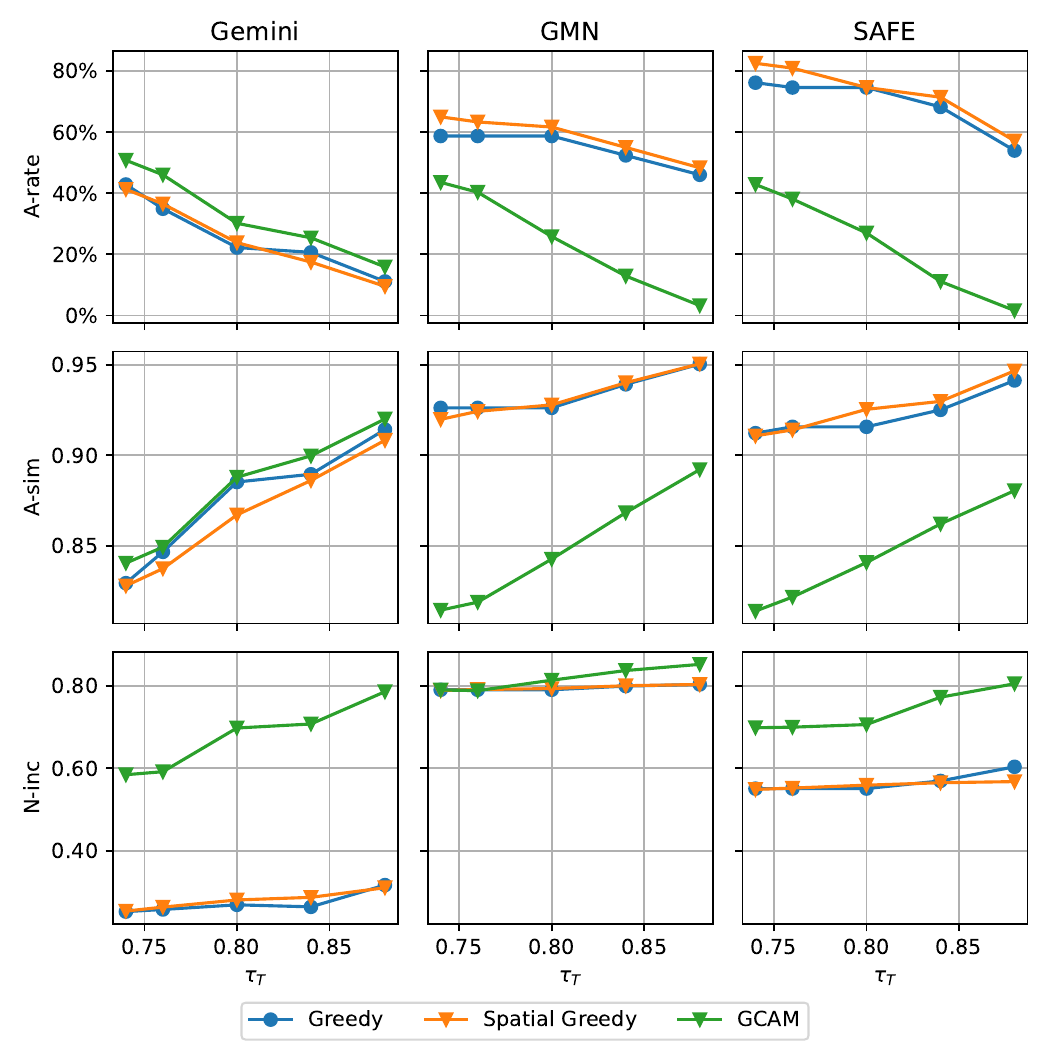}
\caption{Experiments on the three models subject of \textbf{black-box} and \textbf{white-box} attackers in the \textbf{targeted} scenario, on the \textsf{Mirai} dataset for a different success threshold $\mT \in (0.74,0.88)$ in setting \textbf{C4}. In case of \textbf{black-box} attacker, we test the Spatial Greedy approach against the target models with $\varepsilon = 0.1$, $r = 0.75$, $|\texttt{CAND}| = 400$. In case of \textbf{white-box} attacker, we test GCAM attack with 20k iterations against GEMINI, with 1k iterations against GMN, and with 1k iterations against SAFE.} \label{img:miraiTar}
\end{figure}

Figure \ref{img:miraiTar} reports on our targeted attacks, comparing Greedy, Spatial Greedy, and the white-box GCAM for the metrics of \textbf{A-rate}, \textbf{A-sim} and \textbf{N-inc}. For brevity, we focus on the performant \textbf{C4} configuration from the main evaluation.

For the \textbf{A-rate}, when attacking GMN and SAFE, Spatial Greedy has an edge on both Greedy and GCAM, with the latter performing markedly worse than the two black-box ones. With Gemini, Spatial Greedy and Greedy perform similarly, with both resulting below GCAM. This behaviour is consistent with the main evaluation results (cf. Figure~\ref{img:blackWhite_Targeted}).

In more detail, with GMN, the average increase of {A-rate} for Spatial Greedy over Greedy is $3.73$ (max. of $6.27$ at $\mT=0.74$, min. of $2.27$ at $\mT=0.88$).
With SAFE, this increase is $3.81$ (max. of $6.35$ at $\mT=0.74$; min. of zero at $\mT=0.8$). With Gemini, GCAM is the best attack with an average $7.94$ increase over Spatial Greedy (max. of $9.52\%$ at $\mT=0.74$; min. of $6.35\%$ at $\mT=0.88$).
SAFE remains the easiest model to attack also on this dataset.

Regarding \textbf{A-sim} and \textbf{N-inc}, Spatial Greedy and Greedy perform similarly on GMN and SAFE, whereas on Gemini Spatial Greedy is slightly worse than Greedy for \textbf{A-sim} at lower thresholds. The relative performance of GCAM vs. the black-box attacks resembles the trends discussed in the main evaluation (cf. Figure \ref{img:blackWhite_Targeted}).

\begin{figure}[h!]
\centering
\includegraphics[width=3.5in, trim = 0cm 0.6cm 0cm 0cm]{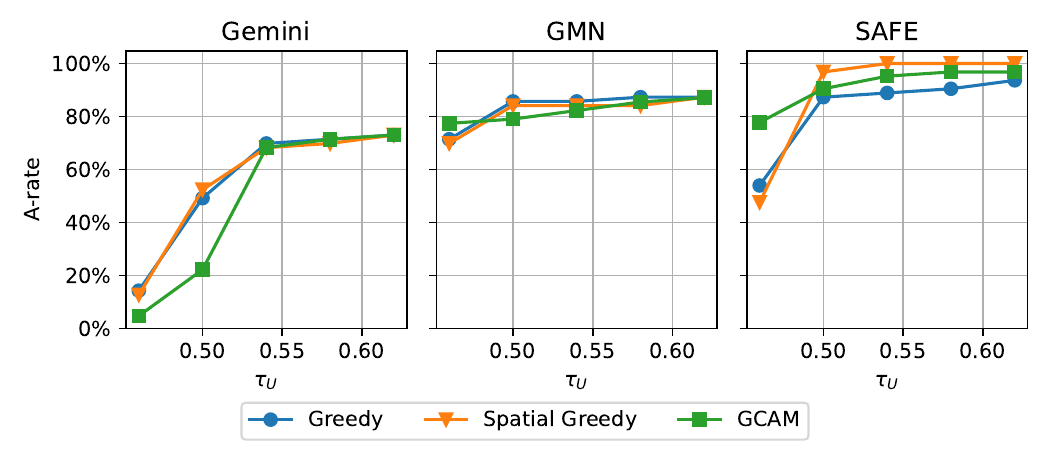}
\caption{Resilience of the three models to \textbf{black-box} and \textbf{white-box} attackers in the \textbf{untargeted} scenario, on the \textsf{Mirai} dataset for a different success threshold \U $\in[0.46, 0.62]$, considering the setting \textbf{C4}. In case of \textbf{black-box} attacker, we test the Spatial Greedy approach against the target models with $\varepsilon = 0.1$, $r = 0.75$, $|\texttt{CAND}| = 400$. In case of \textbf{white-box} attacker, we test GCAM attack with 20k iterations against GEMINI, with 1k iterations against GMN, and with 1k iterations against SAFE.} 
\label{img:miraiUntar}
\end{figure}

Figure~\ref{img:miraiUntar} reports on the experiments we conducted for the untargeted scenario. We note that Spatial Greedy outperforms the other attacks on SAFE (with the exception of GCAM when \U=$0.46$) and performs analogously to them on the other two models. Compared to the main evaluation results, targeted attacks have worse performance than untargeted ones also on this dataset. Moreover, successful targeted attacks continue to require fewer instructions: in particular, across all models, a successful black-box targeted attack needs on average $29.77$ instructions, whereas the untargeted one adds on average $5.27$ instructions.

\section{Related Works}
\label{se:related}
In this section, we first discuss loosely related approaches for attacking image classifiers and natural language processing (NLP) models; then, we describe attacks against source code models. Finally, we discuss prominent attacks against models for binary code analysis.

\subsection{Attacks to Image Classifiers and NLP Models}
Historically, the first adversarial samples targeted image classifiers. The crucial point for these attacks is to introduce inside a clean image instance a perturbation that should not be visible to the human eye while being able to fool the target model, as first pointed out by~\cite{DBLP:journals/corr/SzegedyZSBEGF13} and~\cite{goodfellow2014explaining}.

Most of the attacks modify the original instances using gradient-guided methods. In particular, when computing an adversarial sample, they keep the weights constant while altering the starting input in the direction of the gradient that maximizes (or minimizes, depending on whether the attack is targeted or untargeted) the loss function of the attacked model. The FGSM attack~\cite{goodfellow2014explaining} explicitly implements this technique. Other attacks, such as the Carlini-Wagner~\cite{carlini2017towards} one, generate a noise that is subject to $L_p$-norm constraints to preserve similarity to original objects.

As observed in Section~\ref{sec:invMapping}, adversarial sample generation is possibly easier in the image domain than in the textual one, due to the continuous representation of the original objects.
In the NLP domain, the inputs are discrete objects, a fact that prevents any direct application of gradient-guided methods for adversarial sample generation. Ideally, perturbations to fool deep models for language analysis should be grammatically correct and semantically coherent with the original instance.

One of the earliest methodologies for attacking NLP models is presented in~\cite{DBLP:conf/emnlp/JiaL17}. The authors propose attacks to mislead deep learning-based reading comprehension systems by introducing perturbations in the form of new sentences inside a paragraph, so as to confuse the target model while maintaining intact the original correct answer. The attacks proposed in~\cite{DBLP:conf/naacl/MrksicSTGRSVWY16} and~\cite{DBLP:conf/acl/RenDHC19} focus on finding replacement strategies for words composing the input sequence. Intuitively, valid substitutes should be searched through synonyms; however, this strategy could fall short in considering the context surrounding the word to substitute. Works like~\cite{DBLP:conf/naacl/LiZPCBSD21} and~\cite{DBLP:conf/emnlp/LiMGXQ20} further investigate this idea using BERT-based models for identifying accurate word replacements.

\subsection{Attacks against Models for Source Code Analysis}\label{sec:sc_attacks}
This section covers some prominent attacks against models that work on source code. The techniques from these and other works have limited applicability to binary similarity, as their perturbations may not survive compilation (e.g., variable renaming) or result in marginal differences in compiled code (e.g., turning a while-loop into a for-loop).

The general white-box attack of~\cite{yefet2020adversarial} iteratively substitutes a target variable name in all of its occurrences with an alternative name until a misclassification occurs. 
The attack against plagiarism detection from~\cite{DBLP:journals/pacmpl/Devore-McDonald20} uses genetic programming to augment a program with code lines picked from a pool and validated for program equivalence by checking that an optimizer compiler removes them.
The attack against clone detection from~\cite{zhang2023challenging} combines several semantics-preserving transformations of source code using different optimization heuristic strategies. 

\subsection{Attacks against Models for Binary Code Analysis}\label{sec:related}
We complete our review of related works by covering prominent research on evading ML-based models for analysis of binary code.

Attacks such as~\cite{kolosnjaji2018adversarial,k2018adversarial} to malware detectors based on convolutional neural networks add perturbations in a new non-executable section appended to a Windows PE binary. Both use gradient-guided methods for choosing single-byte perturbations to mislead the model in classifying the whole binary. However, they are ineffective when only actual code is analyzed (e.g., once non-executable sections are stripped).

Pierazzi et al.~\cite{pierazzi2020intriguing} explore transplanting binary code gadgets into a malicious Android program to avoid detection. The attack follows a gradient-guided search strategy based on a greedy optimization. In the initialization phase, they mine from benign binaries code gadgets that modify features that the classifier uses to compute its classification score. In the attack phase, they pick the gadgets that can mostly contribute to the (mis)classification of the currently analyzed malware sample; they insert gadgets in order of decreasing negative contribution, repeating the procedure until misclassification occurs. To preserve program semantics, gadgets are injected into never-executed code portions.

Lucas et al.~\cite{lucas2021malware} target malware classifiers analyzing raw bytes. They propose a functionality-preserving iterative procedure viable for both black-box and white-box attackers. At every iteration, the attack determines a set of applicable transformations for every function in the binary and applies a randomly selected one (following a hill-climbing approach in the black-box scenario or using the gradient in the white-box one). Done via binary rewriting, the transformations are local and include instruction reordering, register renaming, and replacing instructions with equivalent ones of identical length. The results show that these transformations can be effective even against (ML-based) commercial antivirus products, leading the authors to advocate for augmenting such systems with provisions that do not rely on ML. In the context of binary similarity, though, we note that these transformations would have limited efficacy if done on a specific pair of functions: for example, both instruction reordering and register renaming would go completely unnoticed by Gemini and GMN (Section~\ref{sec:gemini} and~\ref{sec:gmn}).

MAB-Malware~\cite{DBLP:conf/asiaccs/SongLAGKY22} is a reinforcement learning-based approach for generating adversarial samples against PE malware classifiers in a black-box context. Adversarial samples are generated through a multi-armed bandit (MAB) model that has to keep the sample in a single, non-evasive state when selecting actions while learning reward probabilities. The goal of the optimization strategy is to maximize the total reward. The set of applicable actions are standard PE manipulation techniques from prior works: header manipulation, section insertion and manipulation (e.g., adding trailing byte), and in-place randomization of an instruction sequence (i.e., replacing it with a semantically equivalent one). Each action is associated with a specific content---a payload---added to the malware when the action is selected. An extensive evaluation is conducted on two popular ML-based classifiers and three commercial antivirus products.

Concurrently to our work, a publicly available technical report proposes FuncFooler~\cite{jia2022funcfooler} as a black-box algorithm for attempting untargeted attacks against ranking systems (i.e., top-$k$ most similar functions) based on binary similarity. The key idea behind the attack is to insert instructions likely to push the source function below the top results returned by the search engine. Insertion points are fixed: specifically, CFG nodes that dominate the exit points of a function. The algorithm picks the instructions directly from those functions with the least similarity in the pool under analysis, then it compensates for their side effects through additional insertions. We note that the attack definition and scenario fundamentally differ from the ones we use in this paper; also, targeted or white-box strategies are not studied.

\section{Limitations and Future Works} \label{se:limitations}
In this paper, we have seen how adding dead code is a natural and effective way to realize appreciable perturbations for a selection of heterogeneous binary similarity systems.

In Section~\ref{ss:perturbation-selection}, we acknowledged how, in the face of defenders that pre-process code with static analysis, our implementation would be limited from having the inserted dead blocks guarded by non-obfuscated branch predicates.

Our experiments suggest that, depending on the characteristics of a given model and pair of functions, the success of an attack may be affected by factors like the initial difference in code size and CFG topology, among others. In this respect, it could be interesting to explore how to alternate our dead-branch addition perturbation, for example, with the insertion of dead fragments within existing blocks.

We believe both limitations could be addressed in future work with implementation effort, whereas the main goal of this paper was to show that adversarial attacks against binary similarity systems are a concrete possibility. For defensive attempts, we could rely on off-the-shelf obfuscation methods (Section~\ref{ss:perturbation-selection}). To enhance our attacks, we could explore more complex patching implementation strategies based on binary rewriting or a modified compiler back-end. Such studies may then include also other performant similarity systems, such as Asm2Vec~\cite{ding2019asm2vec} or jTrans~\cite{jTrans-ISSTA22}.

On a different note, whereas our primary focus has been on the attack side, a compelling research question arises for considering defensive techniques specifically tailored for our scenario. From the extensive literature on adversarial attacks, we can pinpoint several potential directions.

One approach would be to identify appropriate preprocessing techniques that can mitigate our attacks. For instance, one could consider randomly discarding a certain percentage of instructions. Another potential approach involves the design of feature mapping techniques resilient to our attacks; in Section~\ref{ssec:rq2}, we saw that existing solutions already partially resist white-box attacks. Alternatively, one could also develop a classifier designed to detect adversarial samples produced by our methodology or similar ones.

In the hope to foster research in the area, the implementations of our attacks can be made available upon request to fellow researchers, whom are invited to reach us via e-mail.

\section{Conclusions}\label{se:conclusions}
We presented the first study on the resilience of code models for binary similarity to black-box and white-box adversarial attacks, covering targeted and untargeted scenarios. Our tests highlight that current state-of-the-art solutions in the field (Gemini, GMN, and SAFE) are not robust to adversarial attacks crafted for misleading binary similarity models. Furthermore, their resilience against untargeted attacks appears significantly lower in our tests. Our black-box Spatial Greedy technique also shows that an instruction-selection strategy guided by a dynamic exploration of the entire ISA is more effective than using a fixed set of instructions. We hope to encourage follow-up studies by the community to improve the robustness and performance of these systems.

\section*{Acknowledgments}

This work has been carried out while Gianluca Capozzi was enrolled in the Italian National Doctorate on Artificial Intelligence run by Sapienza University of Rome.
This work has been partially supported by project SERICS (PE00000014) under the MUR National Recovery and Resilience Plan funded by the European Union - NextGenerationEU; EU-GUARDIAN project; and Sapienza Ateneo projects (RM1221816C1760BF and AR1221816C754C33). Part of the computational resources have been offered by the AWS Cloud Credit program.

\bibliographystyle{plain}
\bibliography{bibliography.bib}

\end{document}